\documentclass[12pt]{article}
\pdfoutput=1

\usepackage[a4paper,text={16.8cm,22.4cm}]{geometry}
\usepackage{amsmath,amsfonts,braket,slashed,amssymb,tikz,bm,psfrag,graphicx,color,dsfont,euscript}
\usepackage{mathtools}
\usepackage{multicol}
\usepackage{ctable}
\usepackage{amsfonts}
\usepackage{amssymb}
\usepackage{graphicx}
\usepackage{xcolor}
\usepackage{comment}
\usepackage{slashed}
\usepackage{amsmath}
\usepackage{tikz}
\usepackage[small,labelfont=bf]{caption}

\RequirePackage[sort&compress,square,comma,numbers]{natbib}
\allowdisplaybreaks
\begin{document}

\begin{titlepage}

\begin{flushright}
\normalsize
MITP/20-071\\
ZH-TH-48/20\\
November 15, 2022 (v2)
\end{flushright}
\vspace{1.0cm}
\begin{center}
\Large\bf
Effective Field Theory for Leptoquarks
\end{center}

\vspace{0.5cm}
\begin{center}
Bianka Me\c caj$^a$ and Matthias Neubert$^{a,b,c}$\\
\vspace{0.7cm} 
{\sl ${}^a$PRISMA$^+$ Cluster of Excellence \& Mainz Institute for Theoretical Physics\\
Johannes Gutenberg University, 55099 Mainz, Germany\\[3mm]
${}^b$Department of Physics \& LEPP, Cornell University, Ithaca, NY 14853, U.S.A.\\[3mm]
${}^c$Physik-Institut, Universit\"at Z\"urich, CH-8057, Switzerland}
\end{center}
\vspace{0.8cm}
\begin{abstract}
Leptoquarks enter in several extensions of the Standard Model as possible solutions to a number of observed anomalies. We work within the soft-collinear effective theory framework to present a detailed analysis of the decay rates of the three leptoquarks that appear the most in literature, the scalars $S_1$ and $S_3$ and the vector $U_1^\mu$. Using renormalization-group methods we resum the large logarithms arising from the evolution of the Wilson coefficients between the New Physics scale and the electroweak scale. We also derive the tree-level matching relations for the Wilson coefficients in the effective theory for some specific leptoquark models.
\end{abstract}

\end{titlepage}

\tableofcontents
\newpage

\section{Introduction}

Leptoquarks are hypothetical, color-triplet bosonic particles that couple to both leptons and quarks. They appear in several extensions of the Standard Model (SM)\cite{Buchmuller:1986zs,Leurer:1993ap,Davidson:1993qk,Plehn:1997az,Carpentier:2010ue,Dorsner:2017ufx} and were initially predicted in the Pati--Salam model \cite{Pati:1974yy} and other unified theories \cite{Georgi:1974sy,Georgi:1974yf,Fritzsch:1974nn}. Leptoquark couplings can violate the lepton universality of the SM and introduce generation-changing interactions. In recent years, the observation of the $B$-meson anomalies in the measurements of the ratios $R_{D^{(\ast)}}$ \cite{BaBar:2012obs,BaBar:2013mob,LHCb:2015gmp,Belle:2015qfa,Belle:2016dyj,LHCb:2017smo,LHCb:2017rln} and $R_{K^{(\ast)}}$ \cite{LHCb:2014vgu,LHCb:2017avl,LHCb:2019hip,LHCb:2021trn,LHCb:2021lvy} have raised an interest in both vector (spin-1) and scalar (spin-0) leptoquarks. Indeed, the scalar leptoquarks $S_1$ and $S_3$ or the vector leptoquark $U_1$ are widely considered as the most promising candidates which could explain the observed deviations from the SM \cite{Sakaki:2013bfa,Hiller:2014yaa,Buras:2014fpa,Gripaios:2014tna,deMedeirosVarzielas:2015yxm,Bauer:2015knc,Barbieri:2015yvd,Duraisamy:2016gsd,Cox:2016epl,Crivellin:2017zlb,Hiller:2017bzc,Cai:2017wry,Buttazzo:2017ixm,DiLuzio:2017vat,Calibbi:2017qbu,Bordone:2017bld,Smirnov:2018ske,Blanke:2018sro,Greljo:2018tuh,Bordone:2018nbg,Angelescu:2018tyl,Cornella:2019hct,Bigaran:2019bqv,Cornella:2021sby}. Other authors have used leptoquarks as a possible solution to the long-standing problem of the $(g-2)_\mu$ anomaly \cite{Chakraverty:2001yg,Cheung:2001ip,Bauer:2015knc}. 

Current searches for leptoquark pair production at the LHC at $\sqrt{s}=13$~TeV have set lower mass limits in the range between about 1.0 and 1.8~TeV \cite{CMS:2018qqq,ATLAS:2020dsk,ATLAS:2020xov}, depending on model assumptions. Considering also the future high-luminosity upgrade of the LHC, it appears likely that, if such leptoquarks exist, they can be produced on-shell in the coming years. Then the next step would be to study their properties in a model-independent way using an effective field theory (EFT) framework. This is a non-trivial task, however, because the decaying heavy leptoquark cannot be ``integrated out'' entirely in the classical EFT sense. What can be integrated out are its hard quantum fluctuations. Soft-collinear effective theory (SCET) offers a consistent framework to describe the decays of a very heavy particle into highly energetic light degrees of freedom \cite{Bauer:2000yr,Bauer:2001yt,Bauer:2002nz,Beneke:2002ph} (for a review, see \cite{Becher:2014oda}). SCET is a non-local EFT, which was initially developed to study the decays of $B$ mesons into light particles. The approach of using SCET to analyze the decays of heavy particles beyond the SM was initially introduced in \cite{Alte:2018nbn} for the case of a heavy singlet and later applied to models featuring heavy vector-like quarks \cite{Alte:2019iug} and $Z'$ bosons \cite{Heiles:2020plj}. 

In this work we use the SCET formalism to construct the effective Lagrangians that describe the decays of the leptoquark $S_1$, $S_3$, and $U_1$ into jets of SM particles. To make the discussion more interesting, we also allow for the existence of a light right-handed neutrino $\nu_R$ in the particle spectrum, which is a singlet under the SM. We thus add the following terms to the SM Lagrangian: 
\begin{equation}\label{eq:1}
   {\cal L}_{\rm SM}\to {\cal L}_{\rm SM} 
    + \bar\nu_R\,i\slashed{\partial}\,\nu_R - \frac{M_\nu}{2} \left( \bar\nu_R\nu_R^c + \text{h.c.} \right)
    - \big( \bar L_L\bm{Y}_{\nu\,}\tilde\Phi\,\nu_R + \text{h.c.} \big) \,,
\end{equation}
where $\bm{Y}_\nu$ is a new Yukawa matrix, which can give rise to neutrino masses. The presence of the Majorana mass term $M_\nu$ is optional. Here and below, fields carrying a superscript ``$c$'' denote charge-conjugate fields defined as $\psi^c=C\,\bar\psi^T=-i\gamma^2\,\psi^*$, with $C=i\gamma^0\gamma^2$ (in the Weyl representation of the Dirac matrices) being the charge-conjugation matrix .
Such an extension is well motivated, given that leptoquarks often arise in the context of models of neutrino mass generation \cite{Mohapatra:1980qe,Wetterich:1981bx,Buchmuller:1991ce,Buchmuller:1992qc,Khalil:2006yi,Abbas:2007ag}. A consistent EFT description requires charged heavy particles, such as leptoquarks, to be treated within a heavy-particle effective theory framework, similar to the heavy-quark effective theory (HQET) \cite{Eichten:1989zv,Georgi:1990um,Neubert:1993mb}. For each of the three leptoquarks, we construct the operator basis for two-jet decays at leading and next-to-leading order in the EFT power-counting parameter $\lambda\sim v/\Lambda\ll 1$, where $v$ is the scale of electroweak symmetry breaking (which sets the masses of the SM particles) and the New Physics scale $\Lambda\gg v$ is set by the mass of the decaying leptoquark. In addition, we use renormalization-group (RG) techniques to resum the large QCD and electroweak (Sudakov) logarithms in the Wilson coefficients of the leading-power two-jet operators. Throughout our analysis we allow for the existence of generation-changing leptoquark couplings, which is more general than the original Buchm\"uller--R\"uckl--Wyler model \cite{Buchmuller:1986zs}. 

This paper is organized as follows: We begin in Section~\ref{BasicSCET} with a short introduction of the basic elements of SCET relevant for this work. In Section~\ref{HSEFT} we introduce the heavy-particle effective theories needed to describe the soft gauge-boson interactions of charged heavy scalars and vectors. In Section~\ref{sections1} we construct the two-jet operator bases for the leptoquark $S_1$ at leading power ($\sim\lambda^2$) and next-to-leading power leading power ($\sim\lambda^3$), as well as the leading-power operator basis ($\sim\lambda^3$) for decays into three-jet final states. In Sections~\ref{sections3} and \ref{sectionU1} we construct the corresponding two-jet operator bases for the leptoquarks $S_3$ and $U_1$, respectively. In Section~\ref{secwilsoncoeff} we discuss the running of the Wilson coefficients of the leading-order two-jet operators and sum large logarithmic corrections to these coefficients  using RG equations. Lastly, in Section~\ref{sectionmatching} we present the tree-level matching conditions for some concrete leptoquark extensions of the SM. We summarize our main results in Section~\ref{conclusions}. Some technical details of our calculations are collected in an appendix.

\section{Basic elements of SCET}
\label{BasicSCET}

The central idea of SCET lies in identifying the relevant momentum regions for a given process and assigning those momentum regions to quantum fields in the low-energy EFT \cite{Bauer:2000yr,Bauer:2001yt,Bauer:2002nz,Beneke:2002ph}. The relevant momentum regions for the on-shell decays of a heavy particle into energetic light particles are the ``collinear momenta'' carried by the decay products as well as soft momenta, which can be exchanged between these particles. Hard momenta with components of order the mass $M$ of the heavy particle are integrated out in the construction of SCET and are accounted for by the Wilson coefficients of the operators in the low-energy theory. The directions $\vec{n}_i$ of large energy flow in the final state define the so-called ``collinear directions''. For each such direction, and working in the rest frame of the decaying heavy particle, we define two light-like reference vectors $n_i^\mu=\lbrace 1,\vec{n}_i\rbrace$ and $\bar n_i^\mu=\lbrace 1,-\vec{n}_i\rbrace$ (with $n_i^2=\bar n_i^2=0$), such that $n_i\cdot\bar n_i=2$. The freedom to rescale and realign these reference vectors leads to the so-called reparameterization invariance of the SCET Lagrangian \cite{Manohar:2002fd}, which is a remnant of Lorentz invariance.  

The 4-momentum $p_i^\mu$ of a particle can be decomposed in light-cone coordinates such that
\begin{equation}
   p_i^\mu = p_i\cdot\bar n_i\,\frac{n_i^\mu}{2} + p_i\cdot n_i\,\frac{\bar n_i^\mu}{2} + p_{i\perp}^\mu \,.
\end{equation}
For a collinear particles in a jet with direction $\vec{n}_i$, the components scale as $(p_i\cdot n_i, p_i\cdot\bar n_i, p_{i\perp})\sim M(\lambda^2,1,\lambda)$, where $M$ is the mass of the decaying particle and $\lambda\ll 1$ is the expansion parameter of SCET. The components of a soft momentum are such that they all vanish in the limit where $\lambda\to 0$. The exact $\lambda$ scaling depends on the specific process, but in most cases they are either soft, with $k_s\sim M(\lambda,\lambda,\lambda)$, or ultra-soft, with $k_{us}\sim M(\lambda^2,\lambda^2,\lambda^2)$. In this work it will not be necessary to differentiate between these two cases and we will generically refer to soft modes.

Acting with an arbitrary number of $\bar n_i\cdot\partial$ derivatives on an $n_i$-collinear field leaves its $\lambda$ scaling unchanged, since $\bar n_i\cdot p\sim M$ for such a particle. To account for the effect of such derivatives, operators built out of collinear fields need to be allowed to be non-local along the light-like direction $\bar n_i$, for instance 
\begin{equation}\label{collinearfield}
   \psi_{n_i}(x+t\bar n_i) = \sum_{k=0}^\infty\,\frac{t^k}{k!}\,(\bar n_i\cdot\partial)^k\,\psi_{n_i}(x) \,,
\end{equation}
where $t$ is a displacement in the anti-collinear direction. In the effective Lagrangian, operators built out of such collinear fields always appear multiplied by Wilson coefficients that also depend on the $t$ parameters, and these products are integrated over the variables $t$. In this way an arbitrary dependence on the large derivatives $\bar n_i\cdot\partial$ can be accounted for. This property of collinear fields makes SCET a non-local EFT. In order to maintain gauge invariance, it is therefore necessary to introduce $n_i$-collinear Wilson lines, which are defined as \cite{Bauer:2002nz}
\begin{equation}\label{Wilsonline}
   W_{n_i}^{(A)}(x) 
   = P\,\exp\left[ i g_A\,t_A^a\,\int_{-\infty}^0\!ds\,\bar n_i\cdot A_{n_i}^a(x+s\bar n_i) \right] , 
\end{equation}
where $A_{n_i}^a(x)$ is a collinear gauge field, $t_A^a$ is the corresponding group generator in the representation of the field on which the Wilson line is acting, and $g_A$ is the appropriate gauge coupling. 

The effective SCET fields for $n_i$-collinear fermions, scalars, and gauge bosons have a well-defined scaling with respect to the power-counting parameter $\lambda$. Collinear spinor fields are defined with the help of a projector operator $P_{n_i}=\frac{\slashed{n_i}\slashed{\bar{n_i}}}{4}$ (with $P_{n_i}^2=P_{n_i}$), which projects out the large components of the full spinor field in the high-energy limit. One defines
\begin{equation}\label{fielddef}
   \Psi_{n_i}(x) = \frac{\slashed{n}_i\slashed{\bar{n}}_i}{4}\,W_{n_i}^\dagger(x)\,\psi(x) \sim \lambda \,.
\end{equation}
Here the Wilson line $W_{n_i}$ without a superscript ``$(A)$'' is a product of Wilson lines $W_{n_i}^{(A)}$, one for each gauge group under which the field $\psi(x)$ transforms. The fermion field $\Psi_{n_i}$ obeys the constraint
\begin{equation}
   \slashed{{n}}_i\,\Psi_{n_i}(x) = 0 \,.
\end{equation}
For simplicity of notation, we will denote the $n_i$-collinear SM fermion fields by the names of the corresponding particles supplied with a subscript $n_i$. For the special case of the right-handed neutrino, $\nu_{R,n_i}$, there are no Wilson lines, because this field is a singlet under the SM gauge group. Next, an $n_i$-collinear scalar field in SCET, such as the SM Higgs doublet, is defined such that
\begin{equation}\label{fielddefPhi}
   \Phi_{n_i}(x) = W_{n_i}^\dagger(x)\,\phi(x) \sim \lambda \,.
\end{equation}
One refers to the effective fields $\Psi_{n_i}(x)$ and $ \Phi_{n_i}(x)$ as ``gauge-invariant $n_i$-collinear building blocks'', because they are invariant under $n_i$-collinear gauge transformations. The gauge-invariant building block for an $n_i$-collinear gauge boson is defined as a line integral over the corresponding field strength-tensor sandwiched between two Wilson lines \cite{Bauer:2001yt,Hill:2002vw}, i.e.\
\begin{equation}\label{scetgaugeboson}
   \mathcal{A}_{n_i}^\mu(x) = g_A\,\int_{-\infty}^0\!ds\,\bar n_{i\nu} 
    \left[ W_{n_i}^{(A)\dagger} F_{n_i}^{\nu\mu}\,W_{n_i}^{(A)} \right] (x+s\bar n_i) 
    \sim (\lambda^2,0,\lambda) \,.
\end{equation} 
For an abelian gauge group, such as $U(1)_Y$, this expression simplifies to
\begin{equation}\label{Bgaugeinv}
   \mathcal{B}_{n_i}^\mu(x) = g_B\,\int_{-\infty}^0\!ds\,\bar n_{i\nu}\,B_{n_i}^{\nu\mu}(x+s\bar n_i) \,.
\end{equation}
Note that the definition (\ref{scetgaugeboson}) implies that
\begin{equation}\label{gauge fixing}
   \bar n_i\cdot\mathcal{A}_{n_i} = 0 \,.
\end{equation}
This is important, because generically the component $\bar n_i\cdot A_{n_i}^a\sim M$ would be of leading order in power counting. In (\ref{scetgaugeboson}), the introduction of the Wilson lines is effectively equivalent to choosing the light-cone gauge $\bar n_i\cdot A_{n_i}^a=0$. The remaining components of the $n_i$-collinear gauge field scale like the corresponding components of an $n_i$-collinear momentum, i.e.\
\begin{equation}\label{gaugescaling}
   \mathcal{A}_{n_i,\perp}^\mu \sim \lambda \,, \qquad  
   n_i\cdot\mathcal{A}_{n_i} \sim \lambda^2 \,,
\end{equation}
where $\mathcal{A}_{n_i,\perp}^\mu$ is defined as
\begin{equation}\label{vecperp}
   \mathcal{A}_{n_i,\perp}^\mu = \mathcal{A}_{n_i}^\mu - n_i\cdot\mathcal{A}_{n_i}\,\frac{\bar n_i^\mu}{2} \,.
\end{equation}
The component $n_i\cdot\mathcal{A}_{n_i}$ is power suppressed with respect to $\mathcal{A}_{n_i,\perp}^\mu$, and in fact it can always be eliminated using a field redefinition \cite{Marcantonini:2008qn}. This implies that only the transverse components of a collinear gauge field are needed in the construction of the operators in the EFT. 

In principle, it is necessary to introduce SCET fields also for SM particles carrying soft momenta, and these fields have a well-defined $\lambda$ scaling as well. In this work, though, operators containing soft fields would always contribute at higher order in power counting than the operators we will consider. We therefore do not need to specify the definitions of the soft fields.

\section{Heavy-Particle Effective Theory}
\label{HSEFT}

In our framework of describing the decays of heavy leptoquarks into light SM particles, we are integrating out hard quantum fluctuations at the scale of the leptoquark mass $M$. This restricts the interactions of the heavy particles in the low-energy EFT to soft momentum transfer, $|k^\mu|\ll M$, such that the heavy particle remains close to its mass shell under such interactions. The 4-momentum of a heavy particle with mass $M$ can be written as
\begin{equation}
   p^\mu = M v^\mu + k^\mu \,, 
\end{equation} 
where $v^\mu$ is a time-like reference vector (with $v^2=1$), which is identified with the 4-velocity of the heavy particle, and $k^\mu$ is referred to as the ``residual momentum''. In the rest frame of the heavy particle we have $v^\mu=(1,0,0,0)$. 

In this kinematic setup, the soft interactions of an initial-state heavy scalar $S(x)$ can be described by a low-energy EFT built with an effective field $S_v(x)$ defined through the field redefinition
\begin{equation}\label{Svdef}
   S(x) = e^{-i M_S\,v\cdot x}\,S_v(x) \,,   
\end{equation}
where $M_S$ is the mass of the heavy scalar, and the new field carries the residual momentum $k$. Inserting this field redefinition in the Lagrangian of a complex scalar field,
\begin{equation}\label{Lscalar}
   \mathcal{L} = (D^\mu S)^\dagger (D_\mu S) - M_S^2\,S^\dagger S \,,   
\end{equation}
leads to the Lagrangian of heavy-scalar effective theory (HSET). We find
\begin{equation}\label{LHSET}
   \mathcal{L}_{\text{HSET}} = 2 M_S \left[ S_v^\dagger\,i v\cdot D\,S_v
    - \frac{1}{2 M_S}\,S_v^\dagger\,D^2 S_v \right] .
\end{equation}
The second term inside the brackets is suppressed by $k/M_S$ relative to the first one, because a covariant derivative acting on the new field $S_v$ scales like the residual momentum. At leading power, one obtains
\begin{equation}\label{LHSETfinal}
   \mathcal{L}_{\text{HSET}} = 2 M_S \left[ S_v^\dagger\,i v\cdot D\,S_v
    + \mathcal{O}\bigg(\frac{1}{M_S }\bigg) \right] ,
\end{equation} 
which apart from an overall factor $2 M_S$ is the same as the effective Lagrangian of HQET \cite{Georgi:1990um}.\footnote{The prefactor $2M_S$ in the HSET Lagrangian could be eliminated by a rescaling of the field $S_v$. We refrain from doing this, because it would change the canonical dimension of the scalar field.} 
This is natural, because the gauge  interactions of heavy particles exhibit a spin-flavor symmetry in the infinite-mass limit. The mass of the heavy particle then disappears from the effective Lagrangian, and its spin becomes irrelevant. Contrary to HQET, we did not integrate out any degrees of freedom when deriving the Lagrangian (\ref{LHSET}), but we only applied a simple field redefinition. As a result, this Lagrangian is exactly equivalent to the original Lagrangian (\ref{Lscalar}). It does not receive higher-order power corrections, and its operators are not renormalized. In our analysis for the scalar leptoquarks $S_1$ and $S_3$ we will use the Lagrangian in (\ref{LHSETfinal}) to describe the soft gauge-boson exchanges between the leptoquark and the final-state particles.

In a similar fashion, a consistent description of the decay rates of a charged heavy vector particle $V^\mu$ requires a heavy-vector effective theory (HVET), in which one identifies the leading components of the vector field in the infinite-mass limit \cite{Heiles:2020plj}. We start by separating the transverse and longitudinal components of the vector $V^\mu$ using the projection operators
\begin{equation}
   P_{\perp_v\,\nu}^\mu = g_{~\nu}^\mu - v^\mu\,v_\nu \,, \qquad 
   P_{\parallel\,\nu}^\mu = v^\mu\,v_\nu \,. 
\end{equation}
This leads to
\begin{equation}\label{U1perpdef}
   V^\mu = P_{\perp_v\,\nu}^\mu\,V^\nu + P_{\parallel\,\nu}^\mu\,V^\nu
   \equiv V_{\perp_v}^\mu + V_\parallel^\mu \,,
\end{equation}
where $V_{\perp_v}^\mu$ is the field component with polarization perpendicular to the heavy-particle momentum, and $V_\parallel^\mu$ is the component with longitudinal polarization. We use the symbol ``$\perp_v$'' instead of ``$\perp$'' to indicate that in the context of heavy-particle EFTs {\em transverse\/} means {\em orthogonal on the 4-velocity $v^\mu$}, which is different from the meaning of {\em transverse\/} in the context of SCET. 

We now derive the HVET Lagrangian starting from the gauge-invariant Lagrangian for a massive, charged vector field 
\begin{equation}\label{KinetictermU1}
   \mathcal{L} = - \frac12 \left( D^\mu V^\nu - D^\nu V^\mu \right)^\dagger
    \left( D_\mu V_\nu - D_\nu V_\mu \right) + M_V^2\,V_\mu^\dagger V^\mu \,.
\end{equation}
We then perform field redefinitions such that 
\begin{equation}\label{upps}
   V_{\perp_v}^\mu = e^{-i\,M_V v\cdot x}\,V_v^\mu(x) \,, \qquad
   V_\parallel^\mu = e^{-i\,M_V v\cdot x}\,v^\mu\,\mathbb{V}_v(x) \,,
\end{equation}
where
\begin{equation}
   v\cdot V_v(x) = 0 \,.
\end{equation}
After some straightforward manipulations, the Lagrangian that describes the interactions of these effective fields can be brought to the form
\begin{equation}\label{LHVETstep1}
\begin{aligned}
   \mathcal{L}
   &= - 2 M_V\,V_v^{\mu\dagger}\,i v\cdot D\,V_{v,\mu} 
    + V_v^{\mu\dagger} D^2\,V_{v,\mu} - V_{v,\mu}^\dagger\,D_{\perp_v}^\nu D_{\perp_v}^\mu V_{v,\nu} \\
   &\quad + \mathbb{V}_v^\dagger \left( M_V^2+D_{\perp_v}^2 \right) \mathbb{V}_v  
    + \left[ \mathbb{V}_v^\dagger\,iD_{\perp_v}^\mu \left( M_V+iv\cdot D \right) V_{v,\mu} + \text{h.c.} \right] ,
\end{aligned}
\end{equation}
where $D_{\perp_v}^\mu=P_{\perp_v\nu}^\mu\,D^\nu$. This form shows that the effective field $V_v^\mu$ describes massless modes, while the field $\mathbb{V}_v$ describes modes with mass $M_V$. This heavy longitudinal component is integrated out in the construction of the HVET. At tree level, this is accomplished by solving its classical equation of motion, which leads to\footnote{Integrating out the heavy field more carefully using functional methods leads to a non-trivial functional determinant proportional to $\ln\det(M_V^2+D_{\perp_v}^2)$ in the effective Lagrangian of the HVET, which does not contain the field $V_v$ and thus is irrelevant to our discussion in this work.} 
\begin{equation}
   \mathbb{V}_v = - \frac{1}{M_V^2+D_{\perp_v}^2}\,iD_{\perp_v}^\mu \left( M_V+iv\cdot D \right) V_{v,\mu} 
   = - \frac{iD_{\perp_v}^\mu}{M_V}\,V_{v,\mu} + \mathcal{O}\bigg( \frac{1}{M_V^2} \bigg) \,.
\end{equation}
Inserting this solution back into (\ref{LHVETstep1}), we obtain the HVET Lagrangian 
\begin{equation}
\begin{aligned}
   \mathcal{L}_{\text{HVET}}
   &= - 2 M_V\,V_v^{\mu\dagger}\,i v\cdot D\,V_{v,\mu} 
    + V_v^{\mu\dagger} D^2\,V_{v,\mu} - V_{v,\mu}^\dagger\,D_{\perp_v}^\nu D_{\perp_v}^\mu V_{v,\nu} \\
   &\quad - V_{v,\mu}^\dagger \left( M_V+iv\cdot D \right) iD_{\perp_v}^\mu\,
    \frac{1}{M_V^2+D_{\perp_v}^2}\,iD_{\perp_v}^\nu \left( M_V+iv\cdot D \right) V_{v,\nu} \,.
\end{aligned}
\end{equation}
Expanding this expression in powers of $1/M_V$ generates an infinite sets of interaction terms. Since we have integrated out the longitudinal field component $V_\parallel^\mu$ when constructing the HVET, the coefficients of these interactions must in general be renormalized in a non-trivial way. Including only the first-order power corrections, we find
\begin{equation}
   \mathcal{L}_{\text{HVET}}
   = 2 M_V \left[ - V_v^{\mu\dagger}\,i v\cdot D\,V_{v,\mu} 
    + \frac{1}{2 M_V} \left( V_v^{\mu\dagger} D^2\,V_{v,\mu}
    + V_{v,\mu}^\dagger\,[D_{\perp_v}^\mu,D_{\perp_v}^\nu]\,V_{v,\nu} \right) 
    + \mathcal{O}\bigg( \frac{1}{M_V^2} \bigg) \right] .
\end{equation}
In the rest frame (RF) of the heavy meson, the field $V_v^\mu$ only has non-zero spatial components $V_v^k$ with $k=1,2,3$, for which the effective Lagrangian reads
\begin{equation}
   \mathcal{L}_{\text{HVET}} \big|_{\text{RF}}
   = 2 M_V \left[ V_v^{k\dagger}\,i v\cdot D\,V_v^k
    - \frac{1}{2 M_V}\,V_v^{k\dagger} D^2\,V_v^k
    - \frac{1}{2 M_V}\,V_v^{j\dagger}\,[D_{\perp_v}^j,D_{\perp_v}^k]\,V_v^k 
    + \mathcal{O}\bigg( \frac{1}{M_V^2} \bigg) \right] .
\end{equation}
The first two terms are the same as in the scalar case, see (\ref{LHSET}), while the third term accounts for the magnetic interactions between the vector field and the gauge fields. In the infinite-mass limit we recover once again the familiar HQET Lagrangian. In our analysis for the vector leptoquark $U_1$ we will use the leading term in this Lagrangian to describe the soft gauge-boson exchanges between the leptoquark and the final-state particles.

\section{\boldmath SCET formalism for the scalar  leptoquark $S_1(3,1,-\frac{1}{3})$}
\label{sections1}

The scalar leptoquark $S_1$ is a color triplet, an $SU(2)_L$ singlet and has hypercharge $Y=-\frac13$. It thus shares the transformation properties of a right-handed down quark. This particular leptoquark has been studied as a viable solution both to the flavor anomalies and the $(g-2)_\mu$ anomaly \cite{Bauer:2015knc}. In principle, its quantum numbers allow the $S_1$ to couple to operators which can induce proton decay. In the literature, such operators are usually avoided by assuming the realization of certain symmetries, such as Peccei--Quinn symmetry or other discrete symmetries \cite{Bajc:2005zf,Cox:2016epl}.     

\subsection[Leading-power two-jet operators for $S_1$]{\boldmath Leading-power two-jet operators for $S_1$}\label{Sec1}

We start with the SCET Lagrangian describing the decays of the $S_1$ at leading power in the expansion parameter $\lambda$. The decay products are described by the gauge-invariant collinear building blocks discussed in Section~\ref{BasicSCET}, and the leptoquark itself is described within the HSET by the field $S_{1v}$. At lowest order in $\lambda$, the gauge symmetries allow the $S_1$ to couple to two collinear fermions (one of them colored) moving in opposite directions. Hence, the leading-order operators are of $\mathcal{O}(\lambda^2)$. Here and below we use a subscript $n_i$ to denote an $n_i$-collinear field in SCET. Without loss of generality, we choose the final-state quark to move in the direction $n_1$ and the final-state lepton to move in the direction $n_2\simeq\bar n_1$. The most general SCET Lagrangian at $\mathcal{O}(\lambda^2)$ that respects gauge and Lorentz invariance reads
\begin{equation}\label{Lagrangian1}
\begin{aligned}
   \mathcal{L}_{S_1}^{(\lambda^2)}
   &= C_{\bar u_R^c\ell_R S_1^\ast}^{ij}\,\bar u_{R,n_1}^{c,i} \ell_{R,n_2}^j S_{1v}^\ast
    + C_{\bar Q_L^c L_L S_1^\ast}^{ij}\,\bar Q_{L,n_1}^{c,i} i\sigma_2 L_{L,n_2}^j S_{1v}^\ast \\
   &\quad + C_{\bar d_R^c\nu_R S_1^\ast}^{ij}\,\bar d_{R,n_1}^{c,i} \nu_{R,n_2}^j S_{1v}^\ast + \text{ h.c.} \,.
\end{aligned}
\end{equation}
We label the operators and their Wilson coefficients by their field content. The fields $Q_{L,n_1}$ and $L_{L,n_2}$ represent the collinear quark and lepton doublets, while $u_{R,n_1}$ and $d_{R,n_1}$, $\ell_{R,n_2}$, and $\nu_{R,n_2}$ stand for up- and down-type collinear quarks, right-handed collinear lepton, and right-handed collinear neutrino, respectively. The indices  $i,j\in\lbrace 1,2,3\rbrace$ label the fermion families. As mentioned before, we are considering the most general case where a leptoquark can decay into a quark and a lepton of different generations. Once again, fields carrying a superscript ``$c$'' denote charge-conjugate fields. As a result, all the operators in (\ref{Lagrangian1}) violate fermion number conservation by $\Delta F=2$ units. 

In general, as explained in Section~\ref{BasicSCET}, the different collinear fields in the operators in (\ref{Lagrangian1}) can live at different positions, and their Wilson coefficients can depend on the corresponding displacement variables. Thus, considering e.g.\ the first term in the effective Lagrangian more carefully, we should replace
\begin{equation}\label{Wcoeffintegralnew1}
   C_{\bar u_R^c\ell_R S_1^\ast}^{ij}\,\bar u_{R,n_1}^{c,i} \ell_{R,n_2}^j S_{1v}^\ast
   \to \int\!dt_1\,dt_2\,\bar{C}_{\bar u_R^c\ell_R S_1^\ast}^{ij}(\Lambda,t_1,t_2,\mu)
   \left[ \bar u_{R,n_1}^{c,i}(x+t_1\bar n_1)\,\ell_{R,n_2}^j(x+t_2\bar n_2)\,S_{1v}^\ast(x) \right]_\mu \!,
\end{equation}
where we have also indicated the dependence of the (position-dependent) Wilson coefficient and the composite operator on the factorization scale $\mu$. Here $\Lambda\gtrsim M_{S_1}$ represents an ultra-violet (UV) cutoff, and all short-distance contributions above that scale are encoded in the Wilson coefficient functions. Applying a translation transformation to the fermion fields under the integral, then above expression takes the form
\begin{equation}
\begin{aligned}
   &\int\!dt_1\,dt_2\,\bar{C}_{\bar u_R^c\ell_R S_1^\ast}^{ij}(\Lambda,t_1,t_2,\mu)\,
    e^{i t_1\bar n_1\cdot\mathcal{P}_1}\,e^{i t_2\bar n_2\cdot\mathcal{P}_2}
    \left[ \bar u_{R,n_1}^{c,i}(x)\,\ell_{R,n_2}^j(x)\,S_{1v}^\ast(x) \right]_\mu \\
   &\equiv C_{\bar u_R^c\ell_R S_1^\ast}^{ij}(\Lambda,\bar n_1\cdot\mathcal{P}_1,\bar n_2\cdot\mathcal{P}_2,\mu)\,
    \left[ \bar u_{R,n_1}^{c,i} \ell_{R,n_2}^j S_{1v}^\ast \right](x,\mu) \,,
\end{aligned}
\end{equation}
where the coefficient $C$ is the double Fourier transform of $\bar C$. When acting on a product of collinear fields, the momentum operators $\bar n_i\cdot\mathcal{P}_i$ project out the large component of the total $n_i$-collinear momentum carried by all the fields with the index $n_i$. Reparameterization invariance implies that the coefficient $C_{S_1^\ast\bar u_R^c\ell_R}^{ij}(\Lambda,\bar n_1\cdot\mathcal{P}_1,\bar n_2\cdot\mathcal{P}_2,\mu)$ can depend on these operators only through the Lorentz-invariant combination
\begin{equation}
   \mathcal{P}^2 = \frac{n_1\cdot n_2}{2}\,(\bar n_1\cdot\mathcal{P}_1)\,(\bar n_2\cdot\mathcal{P}_2) 
    + \mathcal{O}(\lambda^2) \,,
\end{equation}
whose eigenvalue is equal to $M_{S_1}^2$. The net effect of these manipulations is that we must replace
\begin{equation}\label{Wilsoncoeff}
   C_{\bar u_R^c\ell_R S_1^\ast}^{ij}\,\bar u_{R,n_1}^{c,i} \ell_{R,n_2}^j S_{1v}^\ast
   \to C_{\bar u_R^c\ell_R S_1^\ast}^{ij}(\Lambda,M_{S_1},\mu)
    \left[ \bar u_{R,n_1}^{c,i} \ell_{R,n_2}^j S_{1v}^\ast \right](x,\mu) \,.
\end{equation}
The same replacement applies for the other two operators in the effective Lagrangian. From now on it is implied that the Wilson coefficients of the two-jet operators are always defined as in (\ref{Wilsoncoeff}). Once we have reduced the answer to local operators, we will drop the spacetime argument $x$ for simplicity. We can write the final form of the effective Lagrangian describing two-body decays of the scalar leptoquark $S_1$ at leading power in the compact form  
\begin{equation}\label{LS1newform}
\begin{aligned}
   \mathcal{L}_{S_1}^{(\lambda^2)}
   &= C_{\bar u_R^c\ell_R S_1^\ast}^{ij}(\Lambda,M_{S_1},\mu)\,\mathcal{O}_{\bar u_R^c\ell_R S_1^\ast}^{ij}(\mu)
    + C_{\bar Q_L^c L_L S_1^\ast}^{ij}(\Lambda,M_{S_1},\mu)\,\mathcal{O}_{\bar Q_L^c L_L S_1^\ast}^{ij}(\mu) \\
   &\quad + C_{\bar d_R^c\nu_R S_1^\ast}^{ij}(\Lambda,M_{S_1},\mu)\,\mathcal{O}_{\bar d_R^c\nu_R S_1^\ast}^{ij}(\mu)
    + \text{h.c.} \,,
\end{aligned}
\end{equation}
where we have defined the local basis operators
\begin{equation}\label{S1operators}
\begin{aligned}
   \mathcal{O}_{\bar u_R^c\ell_R S_1^\ast}^{ij}
   &= \bar u_{R,n_1}^{c,i} \ell_{R,n_2}^j S_{1v}^\ast \,, \\
   \mathcal{O}_{\bar Q_L^c L_L S_1^\ast}^{ij}
   &= \bar Q_{L,n_1}^{c,i} i\sigma_2 L_{L,n_2}^j\,S_{1v}^\ast \,, \\
   \mathcal{O}_{\bar d_R^c\nu_R S_1^\ast}^{ij}
   &= \bar d_{R,n_1}^{c,i} \nu_{R,n_2}^j S_{1v}^\ast \,.
\end{aligned}
\end{equation}
The Lagrangian (\ref{LS1newform}) contains only dimension-4 operators and therefore the Wilson coefficients are dimensionless. From an experimental point of view, the first two operators describe decays into two-jet final state, in which the collinear fermions can be accompanied by collinear emissions of gauge bosons. The third operator corresponds to a mono-jet signature plus missing energy, because the right-handed neutrino is invisible in the detector. 

From the operator basis in (\ref{S1operators}) it is straightforward to calculate the tree-level decay rates of the leptoquark $S_1$. For this purpose, the SM fields and Wilson coefficients need to be transformed from the weak basis to the mass basis. We denote the components $\mathrm{C}^{ij}$ of the Wilson coefficients in the mass basis with a straight letter ``C'' rather than the original $C$. The two-body decay rates at $\mathcal{O}(\lambda^2)$ are fixed by kinematics, and in the limit of massless final-state particles we obtain
\begin{equation}\label{S1ratesLO}
\begin{aligned}
   \Gamma(S_1\to u_R^i\ell_R^j)
   &= \frac{M_{S_1}}{16\pi}\,\big| \mathrm{C}_{\bar u_R^c\ell_R S_1^\ast}^{ij} \big|^2 \,, \\
   \Gamma(S_1\to u_L^i\ell_L^j)
   &= \frac{M_{S_1}}{16\pi}\,\big| \mathrm{C}_{\bar Q_L^c L_L S_1^\ast}^{ij} \big|^2 \,, \\
   \Gamma(S_1\to d_L^i\nu_L^j)
   &= \frac{M_{S_1}}{16\pi}\,\big| \mathrm{C}_{\bar Q_L^c L_L S_1^\ast}^{ij} \big|^2 \,, \\
   \Gamma(S_1\to d_R^i\nu_R^j)
   &= \frac{M_{S_1}}{16\pi}\,\big| \mathrm{C}_{\bar d_R^c\nu_R S_1^\ast}^{ij} \big|^2 \,.
\end{aligned}
\end{equation}
For different rates only differ in their Wilson coefficients, and the decays $S_1\to u_L^i\ell_L^j$ and $S_1\to d_L^i\nu_L^j$ have the same rate due to $SU(2)_L$ symmetry.

\subsection[Subleading-power two-jet operators for $S_1$]{\boldmath Subleading-power two-jet operators for $S_1$}
\label{secs1sub}

It is of interest to further explore the SCET Lagrangian beyond the leading power, as this will give access to different decay modes. At $\mathcal{O}(\lambda^3)$, the leptoquark $S_1$ can decay into two- and three-jet final states. In both cases the relevant operators contain up to three collinear fields. If two of these three fields belong to the same collinear sector, they share the large component $\bar n_i\cdot p_i$ of the total collinear momentum $p_i$. It follows that one of the two fields carries the large component $u\,\bar n_i\cdot p_i$ and the other one $(1-u)\,\bar n_i\cdot p_i$, where $0<u<1$, since the large components of collinear momenta are always positive. In the definition of the effective Lagrangian one needs to integrate over the value of $u$. Applying the restrictions imposed by gauge and Lorentz invariance, we find that the effective Lagrangian for two-jet decays of the leptoquark $S_1$ at subleading order in power counting is  
\begin{equation}\label{subS1}
\begin{aligned}
   \mathcal{L}_{S_1}^{(\lambda^3)} \Big|_{\text{2 jet}}
   &= \frac{1}{\Lambda}\,\Bigg[
    C_{\bar d_R\tilde\Phi^\dagger L_L S_1}^{(0)\,ij}(\Lambda,M_{S_1},\mu)\,
    \mathcal{O}_{\bar d_R\tilde\Phi^\dagger L_L S_1}^{(0)\,ij}(\mu) \\
   &\hspace{1.1cm} + C_{\bar Q_L\Phi\nu_R S_1}^{(0)\,ij}(\Lambda,M_{S_1},\mu)\,
    \mathcal{O}_{\bar Q_L\Phi\nu_R S_1}^{(0)\,ij}(\mu) \\
   &\hspace{1.1cm} + \sum_{k=1,2}\,\int_0^1\!du\,\Bigg( 
    C_{\bar d_R\tilde\Phi^\dagger L_L S_1}^{(k)\,ij}(\Lambda,M_{S_1},\mu,u)\,
    \mathcal{O}_{\bar d_R\tilde\Phi^\dagger L_L S_1}^{(k)\,ij}(\mu,u) \\ 
   &\hspace{1.7cm} + C_{\bar Q_L\Phi\nu_R S_1}^{(k)\,ij}(\Lambda,M_{S_1},\mu,u)\,
    \mathcal{O}_{\bar Q_L\Phi\nu_R S_1}^{(k)\,ij}(\mu,u) \\
   &\hspace{1.7cm} + C_{\bar d_R B\nu_R S_1}^{(k)\,ij}(\Lambda,M_{S_1},\mu,u)\,
    \mathcal{O}_{\bar d_R B\nu_R S_1}^{(k)\,ij}(\mu,u) \Bigg) + \text{h.c.} \Bigg] \,.
\end{aligned}
\end{equation} 
The operators with a superscript ``(0)'' contain a zero-momentum Higgs doublet $\Phi^{(0)}$, which has the gauge quantum numbers as the Higgs doublet but does not transform under collinear gauge transformations. After electroweak symmetry breaking this field gets replaced by \cite{Alte:2018nbn}
\begin{equation}\label{phizero}
   \Phi^{(0)}\,\stackrel{\textrm{EWSB}}{\longrightarrow}\,\frac{1}{\sqrt{2}}
    \left( \begin{array}{c} 0 \\ v \end{array} \right) .
\end{equation}
These operators are defined as
\begin{equation}\label{Eq.1}
\begin{aligned}
   \mathcal{O}_{\bar d_R\tilde\Phi^\dagger L_L S_1}^{(0)\,ij}
   &= \bar d_{R,n_1}^i \tilde\Phi^{(0)\dagger} L_{L,n_2}^j S_{1v} \,, \\
   \mathcal{O}_{\bar Q_L\Phi\nu_R S_1}^{(0)\,ij}
   &= \bar Q_{L,n_1}^i \Phi^{(0)} \nu_{R,n_2}^j S_{1v} \,,
\end{aligned}
\end{equation} 
where $\tilde\Phi^{(0)}=i\sigma_2\,\Phi^{(0)\ast}$. For the remaining basis operators, the superscript ``$(k)$'' indicates in which collinear direction $n_k$ (with $k=1$ or 2) the third jet is emitted. We define
\begin{equation}\label{Eq.2}
\begin{aligned}
   \mathcal{O}_{\bar d_R\tilde\Phi^\dagger L_L S_1}^{(1)\,ij}(u)
   &= \bar d_{R,n_1}^i \tilde\Phi_{n_1}^{(u)\dagger} L_{L,n_2}^j S_{1v} \,, \\
   \mathcal{O}_{\bar d_R\tilde\Phi^\dagger L_L S_1}^{(2)\,ij}(u)
   &= \bar d_{R,n_1}^i \tilde\Phi_{n_2}^{(u)\dagger} L_{L,n_2}^j S_{1v} \,, \\
   \mathcal{O}_{\bar Q_L\Phi\nu_R S_1}^{(1)\,ij}(u)
   &= \bar Q_{L,n_1}^i \Phi_{n_1}^{(u)} \nu_{R,n_2}^j S_{1v} \,, \\
   \mathcal{O}_{\bar Q_L\Phi\nu_R S_1}^{(2)\,ij}(u)
   &= \bar Q_{L,n_1}^i \Phi_{n_2}^{(u)} \nu_{R,n_2}^j S_{1v} \,, \\
   \mathcal{O}_{\bar d_R B\nu_R S_1}^{(1)\,ij}(u)
   &= \bar d_{R,n_1}^i \slashed{\mathcal{B}}_{n_1}^{\perp(u)} \nu_{R,n_2}^j S_{1v} \,, \\
   \mathcal{O}_{\bar d_R B\nu_R S_1}^{(2)\,ij}(u)
   &= \bar d_{R,n_1}^i \slashed{\mathcal{B}}_{n_2}^{\perp(u)} \nu_{R,n_2}^j S_{1v} \,.
\end{aligned}
\end{equation} 
Note that two fermion fields cannot be emitted in the same $n_i$ direction, since that would give a vanishing contribution due to the projection operators included in the definition (\ref{fielddef}). Since now all basis operators have mass dimension~5, we have extracted a factor $1/\Lambda$ in (\ref{subS1}) to ensure that the Wilson coefficients are dimensionless functions. A superscript ``$(u)$'' indicates that the corresponding $n_i$-collinear field carries the fraction $u$ of the total collinear momentum $p_i$. Explicitly, we have e.g.\ \cite{Alte:2018nbn}
\begin{equation}
  \Phi_{n_1}^{(u)} \equiv \delta\bigg(u-\frac{\bar n_1\cdot\mathcal{P}_\Phi}{\bar n_1\cdot\mathcal{P}_1}\bigg)\,
   \Phi_{n_1} \,,
\end{equation}
where $\mathcal{P}_1$ is the operator for the total $n_1$-collinear momentum. The last two operators contain the gauge-invariant building block for the hypercharge gauge boson, as defined in (\ref{Bgaugeinv}). Explicitly, we have $\slashed{\mathcal{B}}^{\perp}_{n_i}=\gamma^\mu\,\mathcal{B}_{\mu,n_i}^\perp$, with the perpendicular component $\mathcal{B}_{\mu,n_i}^\perp$ defined as in (\ref{vecperp}). The component $n_i\cdot\mathcal{B}_{n_i}\sim\lambda^2$ is further power suppressed and does not enter at this order.

Note that there are no charge-conjugate fields arising in the operator basis at $\mathcal{O}(\lambda^3)$, and therefore all operators conserve fermion number. This implies that there are no interference effects in the decay rates from two-jet operators at leading and subleading order in $\lambda$. 

At tree level, only the two operators in (\ref{Eq.1}) give rise to two-body decay rates of the leptoquark $S_1$. For the relevant decay rates we obtain
\begin{equation}\label{S12bodysuppressed}
\begin{aligned}
   \Gamma(S_1\to d_R^i\bar\nu_L^j)
   &= \frac{v^2}{2\Lambda^2}\,\frac{M_{S_1}}{16\pi}\,
    \big| \mathrm{C}_{\bar d_R\tilde\Phi^\dagger L_L S_1}^{(0)\,ij} \big|^2 \,, \\
   \Gamma(S_1\to d_L^i\bar\nu_R^j)
   &= \frac{v^2}{2\Lambda^2}\,\frac{M_{S_1}}{16\pi}\,\big| \mathrm{C}_{\bar Q_L\Phi\nu_R S_1}^{(0)\,ij} \big|^2 \,.
\end{aligned}
\end{equation}
Both processes give rise to mono-jet signatures in the detector. Their rates are suppressed by a factor of order $v^2/\Lambda^2$ compared to the leading-power two-body decay rates in (\ref{S1ratesLO}).  

\subsection[Leading-power three-jet operators for $S_1$]{\boldmath Leading-power three-jet operators for $S_1$}

At $\mathcal{O}(\lambda^3)$ in power counting, the effective SCET Lagrangian also contains operators describing three-jet decays of the leptoquark $S_1$. They have the same form as the operators in (\ref{Eq.2}), but now with all three fields belonging to different collinear sectors. We thus write the leading-order three-jet Lagrangian in the form
\begin{equation}\label{3jetS1}
\begin{aligned}
   \mathcal{L}_{S_1}^{(\lambda^3)} \Big|_{\text{3 jet}}
   &= \frac{1}{\Lambda}\,\Bigg[ 
    D_{\bar d_R\tilde\Phi^\dagger L_L S_1}^{ij}(\Lambda,M_{S_1},\lbrace m_{kl}^2\rbrace,\mu)\,
    \mathcal{O}_{\bar d_R\tilde\Phi^\dagger L_L S_1}^{ij}(\mu) \\
   &\hspace{1.2cm} + D_{\bar Q_L\Phi\nu_R S_1}^{ij}(\Lambda,M_{S_1},\lbrace m_{kl}^2\rbrace,\mu)\,
    \mathcal{O}_{\bar Q_L\Phi\nu_R S_1}^{ij}(\mu) \\
   &\hspace{1.2cm} + D_{\bar d_R B\nu_R S_1}^{ij}(\Lambda,M_{S_1},\lbrace m_{kl}^2\rbrace,\mu)\,
    \mathcal{O}_{\bar d_R B\nu_R S_1}^{ij}(\mu) + \text{h.c.} \Bigg] \,,
\end{aligned}
\end{equation}
where the basis operators are given by
\begin{equation}\label{ZandgammadecayS1}
\begin{aligned}
   \mathcal{O}_{\bar d_R\tilde\Phi^\dagger L_L S_1}^{ij}
   &= \bar d_{R,n_1}^i \tilde\Phi_{n_3}^\dagger L_{L,n_2}^j S_{1v} \,, \\
   \mathcal{O}_{\bar Q_L\Phi\nu_R S_1}^{ij}
   &= \bar Q_{L,n_1}^i \Phi_{n_3} \nu_{R,n_2}^j S_{1v} \,, \\
   \mathcal{O}_{\bar d_R B\nu_R S_1}^{ij}
   &= \bar d_{R,n_1}^i \slashed{\mathcal{B}}_{n_3}^\perp \nu_{R,n_2}^j S_{1v} \,.
\end{aligned}
\end{equation}
Here $n_1$, $n_2$, $n_3$ are three light-like directions, each defining a jet signature (or a direction of large missing energy) in the experiment. The Wilson coefficients, which we label by $D_{\dots}^{ij}$ in the three-jet case, can depend on the invariant masses squared $m_{kl}^2$ for any pair of final-state particles ($k<l\in\lbrace 1,2,3\rbrace$), in addition to their dependence on the New Physics scale $\Lambda$ and the leptoquark mass $M_{S_1}$ \cite{Alte:2018nbn}. 

As shown in (\ref{fielddefPhi}), the $n_3$-collinear scalar field $\Phi_{n_3}$ is defined by multiplying the SM Higgs doublet with a collinear Wilson line. After electroweak symmetry breaking, this field takes the form
\begin{equation}
   \Phi_{n_3}(0)\,\stackrel{\textrm{EWSB}}{\longrightarrow}\,\frac{1}{\sqrt{2}}\,W_{n_3}^\dagger(0)
    \left(\begin{array}{c} 0 \\ v+h_{n_3}(0) \end{array} \right) .
\end{equation}
The Wilson line can be expressed in terms of the mass eigenstates of the electroweak gauge bosons $W_{n_i}^\pm$, $Z_{n_i}$ and the photon $A_{n_i}$, i.e.\ 
\begin{equation}\label{Phiwilsonline}
   W_{n_3}(0)
   = P\exp \left[ \frac{ig}{2} \int_{-\infty}^0\!ds \left( \begin{array}{ccc}
    \frac{c_w^2-s_w^2}{c_w}\,\bar n_3\cdot Z_{n_3} + 2 s_w\,\bar n_3\cdot A_{n_3} 
    && \sqrt{2}\,\bar n_3\cdot W_{n_3}^+ \\
    \sqrt{2}\,\bar n_3\cdot W_{n_3}^- 
    && - \frac{1}{c_w}\,\bar n_3\cdot Z_{n_3}
    \end{array} \right) (s\bar n_3)\right] ,
\end{equation}
where $c_w\equiv\cos\theta_w$ and $s_w\equiv\sin\theta_w$ are the cosine and sine of the weak mixing angle. The presence of this Wilson line gives rise to additional three-body decays for all the leptoquark operators where $\Phi_{n_3}$ is present, linking final states containing a physical Higgs boson to those involving electroweak gauge bosons.
          
From the three-jet Lagrangian in (\ref{3jetS1}) we can compute the squared matrix elements for the various decay rates, summed over polarizations. For decays into two fermions and a Higgs boson, we find the reparameterization-invariant expressions
\begin{equation}
\begin{aligned}
   \sum_{\rm pol.}\,\big| \mathcal{M}(S_1\to d_R^i\bar\nu_L^j h) \big|^2
   &= \frac{\big| \mathrm{D}_{\bar d_R\tilde\Phi^\dagger L_L S_1}^{ij} \big|^2}{4\Lambda^2}\,
    (n_1\cdot n_2)\,(\bar n_1\cdot p_1)\,(\bar n_2\cdot p_2) \,, \\
   \sum_{\rm pol.}\,\big| \mathcal{M}(S_1\to d_L^i\bar\nu_R^j h) \big|^2
   &= \frac{\big| \mathrm{D}_{\bar Q_L\Phi\nu_R S_1}^{ij} \big|^2}{4\Lambda^2}\,
    (n_1\cdot n_2)\,(\bar{n}_1\cdot p_1)\,(\bar{n}_2\cdot p_2) \,,
\end{aligned}
\end{equation}
where $p_1=p_d$ and $p_2=p_\nu$. We can now use the fact that
\begin{equation}
   \frac{1}{2}\,(n_1\cdot n_2)\,(\bar n_1\cdot p_1)\,(\bar n_2\cdot p_2) 
   = \left( \frac{n_1}{2}\,\bar n_1\cdot p_1 + \frac{n_2}{2}\,\bar n_2\cdot p_2 \right)^2
   = (p_1+p_2)^2 + \mathcal{O}(\lambda^2) \,,
\end{equation}
which up to power corrections is equal to the invariant mass squared of the down-quark--neutrino pair, $m_{d\nu}^2$. The differential decay rates (Dalitz distributions) for the above decay modes are then obtained as
\begin{equation}
\begin{aligned}
   \frac{d^2\Gamma(S_1\to d_R^i\bar\nu_L^j h)}{dm_{d\nu}^2\,dm_{dh}^2}
   &= \frac{1}{512\pi^3}\,\frac{\big| \mathrm{D}_{\bar d_R\tilde\Phi^\dagger L_L S_1}^{ij} \big|^2}{\Lambda^2}\,
    \frac{m_{d\nu}^2}{M_{S_1}^3} \,, \\
   \frac{d^2\Gamma(S_1\to d_L^i\bar\nu_R^j h)}{dm_{d\nu}^2\,dm_{dh}^2}
   &= \frac{1}{512\pi^3}\,\frac{\big| \mathrm{D}_{\bar Q_L\Phi\nu_R S_1}^{ij} \big|^2}{\Lambda^2}\,
    \frac{m_{d\nu}^2}{M_{S_1}^3} \,.
\end{aligned}
\end{equation} 
For simplicity we neglect the masses of all SM particles, which is a reasonable approximation for a leptoquark ,ass at or above the TeV scale. In the massless limit, the phase space boundaries are such that 
\begin{equation}
   0\le m_{dh}^2 + m_{d\nu}^2\le M_{S_1}^2 \,.
\end{equation}

Analogous expressions are obtained for the decay modes in which the Higgs boson is replaced by a longitudinally polarized $W^\pm$ or $Z$ boson. From the structure of (\ref{Phiwilsonline}), we find
\begin{equation}
\begin{aligned}
   \frac{d^2\Gamma(S_1\to d_R^i\bar\nu_L^j Z)}{dm_{d\nu}^2\,dm_{dZ}^2}
   &=\,\frac{1}{512\pi^3}\,\frac{\big| \mathrm{D}_{\bar d_R\tilde\Phi^\dagger L_L S_1}^{ij} \big|^2}{\Lambda^2}\,
    \frac{m_{d\nu}^2}{M_{S_1}^3} \,, \\
   \frac{d^2\Gamma(S_1\to d_R^i\bar\ell_L^j W^-)}{dm_{d\ell}^2\,dm_{dW}^2}
   &= \frac{1}{256\pi^3}\,\frac{\big| \mathrm{D}_{\bar d_R\tilde\Phi^\dagger L_L S_1}^{ij} \big|^2}{\Lambda^2}\,
    \frac{m_{d\ell}^2}{M_{S_1}^3} \,,\\
   \frac{d^2\Gamma(S_1\to d_L^i\bar\nu_R^j Z)}{dm_{d\nu}^2\,dm_{dZ}^2}
   &= \frac{1}{512\pi^3}\,\frac{\big| \mathrm{D}_{\bar Q_L\Phi\nu_R S_1}^{ij} \big|^2}{\Lambda^2}\,
    \frac{m_{d\nu}^2}{M_{S_1}^3} \,,\\
   \frac{d^2\Gamma(S_1\to u_L^i\bar\nu_R^j W^-)}{dm_{u\nu}^2\,dm_{uW}^2}
   &= \frac{1}{256\pi^3}\,\frac{\big| \mathrm{D}_{\bar Q_L\Phi\nu_R S_1}^{ij} \big|^2}{\Lambda^2}\,
    \frac{m_{u\nu}^2}{M_{S_1}^3} \,.
\end{aligned}
\end{equation}

In a similar way, we can compute the differential decay rates mediated by the chirality-preserving operator shown in the third line of (\ref{ZandgammadecayS1}), where the field for the (transversely polarized) hypercharge gauge boson must be expressed in terms of the fields for the physical $Z$ boson and photon. We find
\begin{equation}\label{3bodys12}
\begin{aligned}
   \frac{d^2\Gamma(S_1\to d_R^i \bar\nu_R^j\gamma)}{dm_{d\nu}^2\,dm_{d\gamma}^2}
   &= \frac{\alpha}{32\pi^2}\,\frac{\big| \mathrm{D}_{\bar d_R B\nu_R S_1}^{ij} \big|^2}{\Lambda^2}\,
    \frac{m^2_{d\nu}}{M^3_{S_1}}\,
    \frac{\left(m_{d\gamma}^2\right)^2+\left(m^2_{\nu\gamma}\right)^2}{\left(M^2_{S_1}-m^2_{d\nu}\right)^2} \,, \\
  \frac{d^2\Gamma(S_1\to d_R^i\bar\nu_R^j Z)}{dm_{d\nu}^2\,dm_{dZ}^2}
  &= \frac{\alpha}{32\pi^2}\,\frac{s_w^2}{c_w^2}\,
   \frac{\big| \mathrm{D}_{\bar d_R B\nu_R S_1}^{ij} \big|^2}{\Lambda^2}\,\frac{m_{d\nu}^2}{M^3_{S_1}}\,
   \frac{\left(m_{dZ}^2\right)^2+\left(m^2_{\nu Z}\right)^2}{\left(M^2_{S_1}-m^2_{d\nu}\right)^2} \,,
\end{aligned}
\end{equation} 
where $\alpha$ is the electromagnetic coupling constant. We sum over the two perpendicular polarization vectors of the gauge bosons using
\begin{equation}
   \sum_{i=1,2}\,\epsilon_\perp^\mu(p_3)\,\epsilon_\perp^{\star\nu}(p_3)
   = - \left( g^{\mu\nu} - \frac{n_3^\mu\,\bar n_3^\nu}{2} - \frac{\bar n_3^\mu\,n_3^\nu}{2} \right) .
\end{equation}
The squared amplitudes summed over polarizations are then proportional to the reparameteri\-zation-invariant quantity \cite{Alte:2018nbn}
\begin{equation}
   \frac{(n_1\cdot n_3)\,(n_2\cdot\bar n_3)+(n_2\cdot n_3)\,(n_1\cdot\bar n_3)}{2n_1\cdot n_2}
   = \frac{\left(m_{13}^2\right)^2 + \left(m_{23}^2\right)^2}{\left(M_{S_1}^2-m_{12}^2\right)^2} \,.
\end{equation}

In all the above results, the cases with a neutrino in the final sate are still a three-jet final state in the SCET sense, even though experimentally the neutrinos are not seen in the detector, so the signature is a two-jet event accompanied by large missing energy. 

This concludes our discussion of the two- and three-jet decays of the scalar leptoquark $S_1$. The formalism we have developed can be extended in a straightforward way to the cases of the other two leptoquarks, $S_3$ and $U_1$. For simplicity, we restrict our analysis to the two-jet decays, which in any case are most interesting from a phenomenological point of view.

\section{\boldmath{SCET formalism for the scalar leptoquark $S_3(3,3,-\frac{1}{3})$}}
\label{sections3}

There are several possible extensions of the SM that try to interpret the observed anomalies in $B$-physics systems. Most of these theoretical models that use scalar leptoquarks as a viable explanation contain both the singlet $S_1$ and another scalar leptoquark $S_3$, which transforms as a triplet under $SU(2)_L$ and has hypercharge $-\frac13$ \cite{Buttazzo:2017ixm,Crivellin:2017zlb}. It is therefore of interest to apply our framework to the triplet $S_3$ and compute its tree-level decay rates.       

\subsection[Leading-power two-jet operators for $S_3$]{\boldmath Leading-power two-jet operators for $S_3$}
\label{s3sec}

We start by constructing the leading-order Lagrangian for two-jet decays. Since $S_3$ is an $SU(2)_{L}$ triplet, it should be understood as $S_3\equiv S_3^a\tau^a$ (summed over $a=1,2,3$), where $\tau^a=\sigma_a/2$ are the generators of $SU(2)_L$. As a result, gauge invariance constrains the operator basis a lot more in this case. Indeed, we find only a single operator mediating the decays of the $S_3$ into two energetic SM particles. It is a dimension-4 operator built out of a quark and a lepton doublet and the effective scalar field $S_{3v}$ defined in HSET, see (\ref{Svdef}). The Lagrangian reads        
\begin{equation}\label{Ltripletleading}
   \mathcal{L}_{S_3}^{(\lambda^2)}
   = C_{\bar Q_L^c S_3^\ast L_L}^{ij}(\Lambda,M_{S_3},\mu)\,\mathcal{O}_{\bar Q_L^c S_3^\ast L_L}^{ij}(\mu)
    + \text{h.c} \,,
\end{equation}
with
\begin{equation}\label{LOopS3}
   \mathcal{O}_{\bar Q_L^c S_3^\ast L_L}^{ij}
   = \bar Q_{L,n_1}^{c,i} i\sigma_2\,S_{3v}^\ast L_{L,n_2}^j \,.
\end{equation}
Here $M_{S_3}$ is the leptoquark mass and $i,j$ are flavor indices. The Wilson coefficient is defined in the same way as in equation (\ref{Wilsoncoeff}) and is dimensionless. As for the case of the leptoquark $S_1$, the Lagrangian (\ref{Ltripletleading}) violates fermion number by two units. 

The lowest-order two-body decay rates are governed by the matrix elements of the Lagrangian in (\ref{Ltripletleading}), which allow for decays into a left-handed quark and a left-handed lepton. The triplet $S_{3}$ contains three particles with different electric charges. We can express it in terms of eigenstates of the charge operator, such that
\begin{equation}
   S_3^{2/3} = \frac{S_3^1-i S_3^2}{\sqrt2} \,, \qquad 
   S_3^{-4/3} = \frac{S_3^1+i S_3^2}{\sqrt2} \,, \qquad
   S_3^{-1/3} = S_3^3 \,,
\end{equation}
where the superscript denotes the electric charge of the corresponding particle. Then the two-body decay rates at $\mathcal{O}(\lambda^2)$ for each particle of the triplet evaluate to
\begin{equation}\label{S3LPdecays}
\begin{aligned}
   \Gamma(S_3^{2/3}\to u_L^i\nu_L^j)
   &= \frac{M_{S_3}}{32\pi}\,\big| \mathrm{C}_{\bar Q_L^c S_3^\ast L_L}^{ij} \big|^2 \,, \\
   \Gamma(S_3^{-4/3}\to d_L^i\ell_L^j)
   &= \frac{M_{S_3}}{32\pi}\,\big| \mathrm{C}_{\bar Q_L^c S_3^\ast L_L}^{ij} \big|^2 \,, \\
   \Gamma(S_3^{-1/3}\to u_L^i\ell_L^j)
   &= \frac{M_{S_3}}{64\pi}\,\big| \mathrm{C}_{\bar Q_L^c S_3^\ast L_L}^{ij} \big|^2 \,, \\
   \Gamma(S_3^{-1/3}\to d_L^i\nu_L^j)
   &= \frac{M_{S_3}}{64\pi}\,\big| \mathrm{C}_{\bar Q_L^c S_3^\ast L_L}^{ij} \big|^2 \,.
\end{aligned}
\end{equation}

\subsection[Subleading-power two-jet operators for $S_3$]{\boldmath Subleading-power two-jet operators for $S_3$}

At $\mathcal{O}(\lambda^3)$, the symmetries allow for a larger number of basis operators for two-jet decays. We find six of them containing fermions with mixed chirality and two featuring fermions of the same chirality, such that
\begin{equation}\label{S32jetsub}
\begin{aligned}
   \mathcal{L}_{S_3}^{(\lambda^3)} \Big|_{\text{2 jet}}
   &= \frac{1}{\Lambda}\,\Bigg[
    C_{\bar Q_L S_3\Phi\nu_R}^{(0)\,ij}(\Lambda,M_{S_1},\mu)\,
    \mathcal{O}_{\bar Q_L S_3\Phi\nu_R}^{(0)\,ij}(\mu) \\
   &\hspace{1.1cm} + C_{\bar d_R\tilde\Phi^\dagger S_3 L_L}^{(0)\,ij}(\Lambda,M_{S_1},\mu)\,
    \mathcal{O}_{\bar d_R\tilde\Phi^\dagger S_3 L_L}^{(0)\,ij}(\mu) \\
   &\hspace{1.1cm} + \sum_{k=1,2}\,\int_0^1\!du\,\Bigg( 
    C_{\bar Q_L S_3\Phi\nu_R}^{(k)\,ij}(\Lambda,M_{S_1},\mu,u)\,
    \mathcal{O}_{\bar Q_L S_3\Phi\nu_R}^{(k)\,ij}(\mu,u) \\ 
   &\hspace{1.7cm} + C_{\bar d_R\tilde\Phi^\dagger S_3 L_L}^{(k)\,ij}(\Lambda,M_{S_1},\mu,u)\,
    \mathcal{O}_{\bar d_R\tilde\Phi^\dagger S_3 L_L}^{(k)\,ij}(\mu,u) \\    
   &\hspace{1.7cm} + C_{\bar d_R S_3 W\nu_R}^{(k)\,ij}(\Lambda,M_{S_1},\mu,u)\,
    \mathcal{O}_{\bar d_R S_3 W\nu_R}^{(k)\,ij}(\mu,u) \Bigg) + \text{h.c.} \Bigg] \,.
\end{aligned}
\end{equation} 
All operators in this Lagrangian conserve fermion number. The operators in the first two lines contain the zero-momentum Higgs doublet and are given by
\begin{equation}\label{2bodyS3sub2}
\begin{aligned}
   \mathcal{O}_{\bar Q_L S_3\Phi\nu_R}^{(0)\,ij}
   &= \bar Q_{L,n_1}^i S_{3v}\Phi^{(0)}\nu_{R,n_2}^j \,, \\
   \mathcal{O}_{\bar d_R\tilde\Phi^\dagger S_3 L_L}^{(0)\,ij}
   &= \bar d_{R,n_1}^i\tilde\Phi^{\dagger(0)} S_{3v} L_{L,n_2}^j \,.
\end{aligned}
\end{equation}
The remaining mixed-chirality operators read 
\begin{equation}\label{S3op2jetsubnew}
\begin{aligned}
   \mathcal{O}_{\bar Q_L S_3\Phi\nu_R}^{(1)\,ij}(u)
   &= \bar Q_{L,n_1}^i S_{3v}\Phi_{n_1}^{(u)}\nu_{R,n_2}^j \,, \\
   \mathcal{O}_{\bar Q_L S_3\Phi\nu_R}^{(2)\,ij}(u)
   &= \bar Q_{L,n_1}^i S_{3v}\Phi_{n_2}^{(u)}\nu_{R,n_2}^j \,, \\
   \mathcal{O}_{\bar d_R\tilde\Phi^\dagger S_3 L_L}^{(1)\,ij}(u)
   &= \bar d_{R,n_1}^i\tilde\Phi_{n_1}^{\dagger(u)} S_{3v} L_{L,n_2}^j \,, \\
   \mathcal{O}_{\bar d_R\tilde\Phi^\dagger S_3 L_L}^{(2)\,ij}(u)
   &= \bar d_{R,n_1}^i\tilde\Phi_{n_2}^{\dagger(u)} S_{3v} L_{L,n_2}^j \,.
\end{aligned}
\end{equation}
The same-chirality operators in (\ref{S32jetsub}) contain the perpendicular component of the gauge-invariant collinear building block for the $SU(2)_L$ gauge bosons, such that
\begin{equation}\label{S3opwW}
\begin{aligned}
   \mathcal{O}_{\bar d_R S_3 W\nu_R}^{(1)\,ij}(u)
   &= \bar d_{R,n_1}^i \mathrm{Tr}\big[ S_{3v}\slashed{\mathcal{W}}^{\perp(u)}_{n_1} \big]\,\nu_{R,n_2}^j \,, \\
   \mathcal{O}_{\bar d_R S_3 W\nu_R}^{(2)\,ij}(u)
   &= \bar d_{R,n_1}^i \mathrm{Tr}\big[ S_{3v}\slashed{\mathcal{W}}^{\perp(u)}_{n_2} \big]\,\nu_{R,n_2}^j \,,
\end{aligned}
\end{equation}
where the trace is over $SU(2)_L$ indices. 

Only the operators containing the zero-momentum Higgs doublet (\ref{2bodyS3sub2}) contribute to the power-suppressed two-body decays at tree level. We find the decay rates
\begin{equation}
\begin{aligned}
   \Gamma(S_3^{2/3}\to u_L^i\bar\nu_R^j)
   &= \frac{v^2}{2\Lambda^2}\,\frac{M_{S_3}}{32\pi}\,
    \big| \mathrm{C}_{\bar Q_L S_3\Phi\nu_R}^{(0)\,ij} \big|^2 \,, \\
   \Gamma(S_3^{-1/3}\to d_L^i\bar\nu_R^j)
   &= \frac{v^2}{2\Lambda^2}\,\frac{M_{S_3}}{64\pi}\,
    \big| \mathrm{C}_{\bar Q_L S_3\Phi\nu_R}^{(0)^{ij}} \big|^2 \,,\\
   \Gamma(S_3^{2/3}\to d_R^i\bar\ell_L^j)
   &= \frac{v^2}{2\Lambda^2}\,\frac{M_{S_3}}{32\pi}\,
    \big| \mathrm{C}_{\bar d_R\tilde\Phi^\dagger S_3 L_L}^{(0)\,ij} \big|^2 \,, \\
   \Gamma(S_3^{-1/3}\to d_R^i\bar\nu_L^j)
   &= \frac{v^2}{2\Lambda^2}\,\frac{M_{S_3}}{64\pi}\,
    \big| \mathrm{C}_{\bar d_R\tilde\Phi^\dagger S_3 L_L}^{(0)\,ij} \big|^2 \,.
\end{aligned}
\end{equation}

\section{\boldmath SCET formalism for the vector leptoquark $U_1(3,1,\frac{2}{3})$}
\label{sectionU1}

The vector $U_1$ is another interesting example from the family of leptoquarks, which has been introduced as a solution to departures from the SM in the flavor sector \cite{Barbieri:2015yvd,Buttazzo:2017ixm}. It is a color triplet, $SU(2)_L$ singlet and has hypercharge $\frac23$. In the following section we analyze its decays at leading order in power counting. The soft interactions of the field $U_1$ are described in HVET, as discussed earlier in Section~\ref{HSEFT}. 

\subsection[Leading-power two-jet operators for $U_1$]{\boldmath Leading-power two-jet operators for $U_1$}

The operator basis is built following the same reasoning as for the other two leptoquarks, where we construct all the possible particle combinations that preserve gauge and Lorentz invariance. Also in this case we find non-vanishing operators containing the right-handed neutrino. At leading order in SCET, we find that the Lagrangian contains three operators, such that
\begin{equation}\label{LagrangianU1}
\begin{aligned}
   \mathcal{L}_{U_1}^{(\lambda^2)}
   &= C_{\bar Q_L U_1 L_L}^{ij}(\Lambda,M_{U_1},\mu)\,\mathcal{O}_{\bar Q_L U_1 L_L}^{ij}(\mu)
    + C_{\bar d_R U_1\ell_R}^{ij}(\Lambda,M_{U_1},\mu)\,\mathcal{O}_{\bar d_R  U_1\ell_R}^{ij}(\mu) \\
   &\quad + C_{\bar u_R U_1\nu_R}^{ij}(\Lambda,M_{U_1},\mu)\,\mathcal{O}_{\bar u_R  U_1\nu_R}^{ij}(\mu) 
    + \text{h.c.} \,,
\end{aligned}
\end{equation}
with the dimension-4 operators
\begin{equation}\label{leadingU1operators}
\begin{aligned}
   \mathcal{O}_{\bar Q_L U_1 L_L}^{ij}
   &= \bar Q_{L,n_1}^i\slashed{U}_{\!1v}^\perp\,L_{L,n_2}^j \,, \\
   \mathcal{O}_{\bar d_R U_1\ell_R}^{ij}
   &= \bar d_{R,n_1}^i\slashed{U}_{\!1v}^\perp\,\ell_{R,n_2}^j \,, \\
   \mathcal{O}_{\bar u_R U_1\nu_R}^{ij}
   &= \bar u_{R,n_1}^i\slashed{U}_{\!1v}^\perp\,\nu_{R,n_2}^j \,.
\end{aligned}
\end{equation}
where $\slashed{U}_{\!1v}^\perp=\gamma_\mu^\perp U_{1v}^\mu$ with the effective field $U_{1v}^\mu$ defined in (\ref{upps}). Note that between the projection operators included in the definition of the collinear fermion fields only the perpendicular components (in the SCET sense) of the Dirac matrices $\gamma_\mu$ survive. In other words, the vector leptoquark is transversely polarized with respect to the directions of the final-state particles. At leading order in power counting, the vector leptoquark decays into two fermions of the same chirality. For the unpolarized decay rates, summed (averaged) over final-state (initial-state) polarizations, we obtain
\begin{equation}
\begin{aligned}
   \Gamma(U_1\to u_L^i\bar\nu_L^j)
   &= \frac{M_{U_1}}{24\pi}\,\big| \mathrm{C}_{\bar Q_L U_1 L_L}^{ij} \big|^2 \,, \\
   \Gamma(U_1\to d_L^i\bar\ell_L^J)
   &= \frac{M_{U_1}}{24\pi}\,\big| \mathrm{C}_{\bar Q_L U_1 L_L}^{ij} \big|^2 \,, \\
   \Gamma(U_1\to d_R^i\bar\ell_R^j)
   &= \frac{M_{U_1}}{24\pi}\,\big| \mathrm{C}_{\bar d_R U_1\ell_R}^{ij} \big|^2 \,, \\
   \Gamma(U_1\to u_R^i\bar\nu_R^j)
   &= \frac{M_{U_1}}{24\pi}\,\big| \mathrm{C}_{\bar u_R U_1\nu_R}^{ij} \big|^2 \,.
\end{aligned}
\end{equation}
Contrary to the cases of the two scalar leptoquarks $S_1$ and $S_3$, the leading-power two-jet operators for the vector leptoquark $U_1$ conserve fermion number.

\subsection[Subleading-power two-jet operators for $U_1$]{\boldmath Subleading-power two-jet operators for $U_1$}

At $\mathcal{O}(\lambda^3)$ we find a much larger basis of two-jet operators for the leptoquark $U_1$. Now both the longitudinal and transverse (in the SCET sense) components of the heavy field appear. We will find that operators involving the longitudinal field component conserve fermion number, while those involving the transverse components change fermion number by two units. It is useful to define the reparameterization-invariant quantity \cite{Heiles:2020plj}
\begin{equation}\label{Pidef}
   \Pi^\mu = \frac{(v\cdot n_2)\,n_1^\mu-(v\cdot n_1)\,n_2^\mu}{n_1\cdot n_2} \,,
\end{equation}
where $\Pi^{\mu}\to -\Pi^{\mu}$ under hermitian conjugation \cite{Heiles:2020plj}. This object is orthogonal to the leptoquark momentum, $v\cdot\Pi=0$, but it is {\em longitudinal\/} in the SCET sense, meaning that it lies in the plane spanned by the light-like vectors $n_1$ and $n_2$. In the rest frame of the leptoquark, we simply have $\Pi^\mu=(0,\vec{e}_z)$.

There is a fair number of operators in this case, and for convenience we write the Lagrangian as a sum of two terms
\begin{equation}
   \mathcal{L}_{U_1}^{(\lambda^3)} \Big|_{\text{2 jet}}
   = \mathcal{L}_{U_1}^{(\lambda^3),\,A} + \mathcal{L}_{U_1}^{(\lambda^3),\,\Phi} \,.
\label{LU13jetsnew}
\end{equation} 
The first Lagrangian contains operators built out of two collinear fermion fields and a gauge field, while the second Lagrangian contains operators built out of two collinear fermion fields and a Higgs field. The gauge-boson Lagrangian reads
\begin{equation}\label{U1subLa}
\begin{aligned}
   \mathcal{L}_{U_1}^{(\lambda^3),\,A}
   &= \frac{1}{\Lambda} \sum_{k=1,2} \int_0^1\!du\,\Bigg[
    \sum_{A=G,W,B} C_{\bar Q_L A U_1 L_L}^{(k)\,ij}(\Lambda,M_{U_1},\mu,u)\,
    \mathcal{O}_{\bar Q_L A U_1 L_L}^{(k)\,ij}(\mu,u) \\
   &\hspace{2.95cm} + \sum_{A=G,B} \Bigg( C_{\bar d_R A U_1\ell_R}^{(k)\,ij}(\Lambda,M_{U_1},\mu,u)\,
    \mathcal{O}_{\bar d_R A U_1\ell_R}^{(k)\, ij}(\mu,u) \\
   &\hspace{4.8cm} + C_{\bar u_R A U_1\nu_R}^{(k)\,ij}(\Lambda,M_{U_1},\mu,u)\,
    \mathcal{O}_{\bar u_R A U_1\nu_R}^{(k)\,ij}(\mu,u) \\[3mm]
   &\hspace{4.8cm} + C_{\bar u_R A U_1\nu_R^c}^{(k)\,ij}(\Lambda,M_{U_1},\mu,u)\,
    \mathcal{O}_{\bar u_R A U_1\nu_R^c}^{(k)\,ij}(\mu,u) \\
   &\hspace{4.8cm} + C_{\bar u_R A U_1\nu_R^c}^{\prime\,(k)\,ij}(\Lambda,M_{U_1},\mu,u)\,
    \mathcal{O}_{\bar u_R A U_1\nu_R^c}^{\prime\,(k)\,ij}(\mu,u)        
    \Bigg) + \text{h.c.} \Bigg] \,,
\end{aligned}
\end{equation}
where the operators are defined as 
\begin{equation}\label{1stoperatorsetU1}
\begin{aligned}
   \mathcal{O}_{\bar Q_L A U_1 L_L}^{(1)\,ij}(u)
   &=\bar Q_{L,n_1}^i\rlap{\hspace{1mm}/}{\mathcal{A}}_{n_1}^{\perp(u)}\,\Pi\cdot U_{1v}\,L_{L,n_2}^j \,, \\
   \mathcal{O}_{\bar Q_L A U_1 L_L}^{(2)\,ij}(u)
   &=\bar Q_{L,n_1}^i\rlap{\hspace{1mm}/}{\mathcal{A}}_{n_2}^{\perp(u)}\,\Pi\cdot U_{1v}\,L_{L,n_2}^j \,, \\
   \mathcal{O}_{\bar d_R A U_1\ell_R}^{(1)\,ij}(u)
   &= \bar d_{R,n_1}^i\rlap{\hspace{1mm}/}{\mathcal{A}}_{n_1}^{\perp(u)}\,\Pi\cdot U_{1v}\,\ell_{R,n_2}^j \,, \\
   \mathcal{O}_{\bar d_R A U_1\ell_R}^{(2)\,ij}(u)
   &= \bar d_{R,n_1}^i\rlap{\hspace{1mm}/}{\mathcal{A}}_{n_2}^{\perp(u)}\,\Pi\cdot U_{1v}\,\ell_{R,n_2}^j \,, \\
   \mathcal{O}_{\bar u_R A U_1\nu_R}^{(1)\,ij}(u)
   &= \bar u_{R,n_1}^i\rlap{\hspace{1mm}/}{\mathcal{A}}_{n_1}^{\perp(u)}\,\Pi\cdot U_{1v}\,\nu_{R,n_2}^j \,, \\
   \mathcal{O}_{\bar u_R A U_1\nu_R}^{(2)\,ij}(u)
   &= \bar u_{R,n_1}^i\rlap{\hspace{1mm}/}{\mathcal{A}}_{n_2}^{\perp(u)}\,\Pi\cdot U_{1v}\,\nu_{R,n_2}^j \,, \\
   \mathcal{O}_{\bar u_R A U_1\nu_R^c}^{(1)\,ij}(u)
   &= g_{\mu\nu}^\perp \bar u_{R,n_1} \mathcal{A}_{n_1}^{\mu\perp(u)} 
    U_{1v}^\nu\nu_{R,n_2}^{c,j} \,, \\
   \mathcal{O}_{\bar u_R A U_1\nu_R^c}^{(2)\,ij}(u)
   &= g_{\mu\nu}^\perp \bar u_{R,n_1} \mathcal{A}_{n_2}^{\mu\perp(u)} 
    U_{1v}^\nu\nu_{R,n_2}^{c,j} \,, \\
   \mathcal{O}_{\bar u_R A U_1\nu_R^c}^{\prime(1)\,ij}(u)
   &= i\epsilon_{\mu\nu}^\perp \bar u_{R,n_1} \mathcal{A}_{n_1}^{\mu\perp(u)} 
    U_{1v}^\nu\nu_{R,n_2}^{c,j} \,, \\
   \mathcal{O}_{\bar u_R A U_1\nu_R^c}^{\prime(2)\,ij}(u)
   &= i\epsilon_{\mu\nu}^\perp \bar u_{R,n_1} \mathcal{A}_{n_2}^{\mu\perp(u)} 
    U_{1v}^\nu\nu_{R,n_2}^{c,j} \,,
\end{aligned}
\end{equation}
where the tensors 
\begin{equation}
   g_{\mu\nu}^\perp 
   = g_{\mu\nu} - \frac{n_{1\mu} n_{2\nu}+n_{2\mu} n_{1\nu}}{n_1\cdot n_2} \,, \qquad
   \epsilon_{\mu\nu}^\perp 
   = \frac{1}{n_1\cdot n_2}\,\epsilon_{\mu\nu\alpha\beta}\,n_2^\alpha\,n_1^\beta
\end{equation}
are used to contract Lorentz indices in the plane transverse to the directions $n_1$ and $n_2$. For each operator the allowed gauge bosons follow from the charges of the fields. In all cases the leptoquark is transversely polarized with respect to the decay axis. Note that the last four operators violate fermion humber by two units.

The second Lagrangian in (\ref{LU13jetsnew}) has the form 
\begin{equation}\label{LagrangianU1sub}
\begin{aligned}
   \mathcal{L}_{U_1}^{(\lambda^3),\,\Phi}
   &= \frac{1}{\Lambda}\,\Bigg[ C_{\bar Q_L \Phi U_1\ell_R}^{(0)\,ij}\,(\Lambda,M_{U_1},\mu)\,
    \mathcal{O}_{\bar Q_L\Phi U_1\ell_R}^{(0)\,ij}(\mu) \\
   &\hspace{1.2cm} + C_{\bar Q_L\tilde\Phi U_1\nu_R}^{(0)\,ij}(\Lambda,M_{U_1},\mu)\,
    \mathcal{O}_{\bar Q_L\tilde\Phi U_1\nu_R}^{(0)\,ij}(\mu) \\[2mm]
   &\hspace{1.2cm} + C^{(0)\,ij}_{\bar d_R\Phi^\dagger U_1 L_L}(\Lambda,M_{U_1},\mu)\,
    \mathcal{O}_{\bar d_R\Phi^\dagger U_1 L_L}^{(0)\,ij}(\mu) \\[2mm]
   &\hspace{1.2cm} + C_{\bar u_R\tilde\Phi^\dagger U_1 L_L}^{(0)\,ij}(\Lambda,M_{U_1},\mu)\,
    \mathcal{O}_{\bar u_R\tilde\Phi^\dagger U_1 L_L}^{(0)\,ij}(\mu) \\[2mm]
   &\hspace{1.2cm} + C_{\bar Q_L\tilde\Phi U_1\nu_R^c}^{(0)\,ij}(\Lambda,M_{U_1},\mu)\,
    \mathcal{O}_{\bar Q_L\tilde\Phi U_1\nu_R^c}^{(0)\,ij}(\mu) \\[2mm]
   &\hspace{1.2cm} + C_{\bar u_R\Phi^\dagger U_1 L_L^c}^{(0)\,ij}(\Lambda,M_{U_1},\mu)\,
    \mathcal{O}_{\bar u_R\tilde\Phi U_1 L_L^c}^{(0)\,ij}(\mu) \\
   &\hspace{1.2cm} + \sum_{k=1,2}\,\int_0^1\!du\,\Bigg( 
    C_{\bar Q_L\Phi U_1\ell_R}^{(k)\,ij}(\Lambda,M_{U_1},\mu,u)\,
    \mathcal{O}_{\bar Q_L\Phi U_1\ell_R}^{(k)\,ij}(\mu,u) \\
   &\hspace{1.85cm} + C_{\bar Q_L\tilde\Phi U_1\nu_R}^{(k)\,ij}(\Lambda,M_{U_1},\mu,u)\,
    \mathcal{O}_{\bar Q_L\tilde\Phi U_1\nu_R}^{(k)\,ij}(\mu,u) \\[2mm]
   &\hspace{1.85cm} + C_{\bar d_R\Phi^\dagger U_1 L_L}^{(k)\,ij}(\Lambda,M_{U_1},\mu,u)\,
    \mathcal{O}_{\bar d_R\Phi^\dagger U_1 L_L}^{(k)\,ij}(\mu,u) \\[2mm]
   &\hspace{1.85cm} + C_{\bar u_R\tilde\Phi^\dagger U_1 L_L}^{(k)\,ij}(\Lambda,M_{U_1},\mu,u)\,
    \mathcal{O}_{\bar u_R\tilde\Phi^\dagger U_1 L_L}^{(k)\,ij}(\mu,u) \\[2mm]
   &\hspace{1.85cm} + C_{\bar Q_L\tilde\Phi U_1\nu_R^c}^{(k)\,ij}(\Lambda,M_{U_1},\mu,u)\,
    \mathcal{O}_{\bar Q_L\tilde\Phi U_1\nu_R^c}^{(k)\,ij}(\mu,u) \\
   &\hspace{1.85cm} + C_{\bar u_R\tilde\Phi U_1 L_L^c}^{(k)\,ij}(\Lambda,M_{U_1},\mu,u)\,
    \mathcal{O}_{\bar u_R\tilde\Phi U_1 L_L^c}^{(k)\,ij}(\mu,u) \Bigg) 
    + \text{h.c.} \Bigg] \,.
\end{aligned}
\end{equation}
In this case, some operators include the perpendicular component of the leptoquark field along with a charge-conjugate fermion field. The operators containing a zero-momentum scalar doublet read
\begin{equation}\label{U1phi0new}
\begin{aligned}
   \mathcal{O}_{\bar Q_L\Phi U_1\ell_R}^{(0)\,ij}
   &= \bar Q_{L,n_1}^i\Phi^{(0)}\,\Pi\cdot U_{1v}\,\ell_{R,n_2}^j \,, \\
   \mathcal{O}_{\bar Q_L\tilde\Phi U_1\nu_R}^{(0)\,ij}
   &=\bar Q_{L,n_1}^i\tilde\Phi^{(0)}\,\Pi\cdot U_{1v}\,\nu_{R,n_2}^j \,, \\
   \mathcal{O}_{\bar d_R\Phi^\dagger U_1 L_L}^{(0)\,ij}
   &= \bar d_{R,n_1}^i\Phi^{\dagger(0)}\,\Pi\cdot U_{1v}\,L_{L,n_2}^j \,, \\
   \mathcal{O}_{\bar u_R\tilde\Phi^\dagger U_1 L_L}^{(0)\,ij}
   &= \bar u_{R,n_1}^i\tilde\Phi^{\dagger(0)}\,\Pi\cdot U_{1v}\,L_{L,n_2}^j \,, \\
   \mathcal{O}_{\bar Q_L\tilde\Phi U_1\nu_R^c}^{(0)\,ij}
   &= \bar Q_{L,n_1}^i\tilde\Phi^{(0)}\slashed{U}_{\!1v}^\perp\,\nu_{R,n_2}^{c,j} \,, \\
   \mathcal{O}_{\bar u_R\Phi^\dagger U_1 L_L^c}^{(0)\,ij}
   &= \bar u_{R,n_1}^i \Phi^{\dagger(0)}\slashed{U}_{\!1v}^\perp\,i\sigma^2 L_{L,n_2}^{c,j} \,.
\end{aligned}
\end{equation}
In the last two operators, which change fermion number by two units, the presence of the charge-conjugate right-handed fields ensures that the product of the chirality projectors does not vanish. The remaining operators are given by 
\begin{equation}\label{U1listsublscalarnew1}
\begin{aligned}
   \mathcal{O}_{\bar Q_L\Phi U_1\ell_R}^{(1)\,ij}
   &= \bar Q_{L,n_1}^i\Phi_{n_1}^{(u)}\,\Pi\cdot U_{1v}\,\ell_{R,n_2}^j \,, \\
   \mathcal{O}_{\bar Q_L\Phi U_1\ell_R}^{(2)\,ij}
   &= \bar Q_{L,n_1}^i\Phi_{n_2}^{(u)}\,\Pi\cdot U_{1v}\,\ell_{R,n_2}^j \,, \\
   \mathcal{O}_{\bar Q_L\tilde\Phi U_1\nu_R}^{(1)\,ij}
   &=\bar Q_{L,n_1}^i\tilde\Phi_{n_1}^{(u)}\,\Pi\cdot U_{1v}\,\nu_{R,n_2}^j \,, \\
   \mathcal{O}_{\bar Q_L\tilde\Phi U_1\nu_R}^{(2)\,ij}
   &=\bar Q_{L,n_1}^i\tilde\Phi_{n_2}^{(u)}\,\Pi\cdot U_{1v}\,\nu_{R,n_2}^j \,, \\
   \mathcal{O}_{\bar d_R\Phi^\dagger U_1 L_L}^{(1)\,ij}
   &= \bar d_{R,n_1}^i\Phi_{n_1}^{\dagger(u)}\,\Pi\cdot U_{1v}\,L_{L,n_2}^j \,, \\
   \mathcal{O}_{\bar d_R\Phi^\dagger U_1 L_L}^{(2)\,ij}
   &= \bar d_{R,n_1}^i\Phi_{n_2}^{\dagger(u)}\,\Pi\cdot U_{1v}\,L_{L,n_2}^j \,, \\
   \mathcal{O}_{\bar u_R\tilde\Phi^\dagger U_1 L_L}^{(1)\,ij}
   &= \bar u_{R,n_1}^i\tilde\Phi_{n_1}^{\dagger(u)}\,\Pi\cdot U_{1v}\,L_{L,n_2}^j \,, \\
   \mathcal{O}_{\bar u_R\tilde\Phi^\dagger U_1 L_L}^{(2)\,ij}
   &= \bar u_{R,n_1}^i\tilde\Phi_{n_2}^{\dagger(u)}\,\Pi\cdot U_{1v}\,L_{L,n_2}^j \,, \\
   \mathcal{O}_{\bar Q_L\tilde\Phi U_1\nu_R^c}^{(1)\,ij}
   &= \bar Q_{L,n_1}^i\tilde\Phi_{n_1}^{(u)}\,\slashed{U}_{\!1v}^\perp\,\nu_{R,n_2}^{c,j} \,, \\
   \mathcal{O}_{\bar Q_L\tilde\Phi U_1\nu_R^c}^{(2)\,ij}
   &= \bar Q_{L,n_1}^i\tilde\Phi_{n_2}^{(u)}\,\slashed{U}_{\!1v}^\perp\,\nu_{R,n_2}^{c,j} \,, \\
   \mathcal{O}_{\bar u_R\Phi^\dagger U_1 L_L^c}^{(1)\,ij}
   &= \bar u_{R,n_1}^i \Phi_{n_1}^{\dagger(u)}\,\slashed{U}_{\!1v}^\perp\,i\sigma^2 L_{L,n_2}^{c,j} \,, \\
   \mathcal{O}_{\bar u_R\Phi^\dagger U_1 L_L^c}^{(2)\,ij}
   &= \bar u_{R,n_1}^i \Phi_{n_2}^{\dagger(u)}\,\slashed{U}_{\!1v}^\perp\,i\sigma^2 L_{L,n_2}^{c,j} \,.
\end{aligned}
\end{equation}

Only the operators containing a zero-momentum Higgs doublet contribute at tree-level to the power-suppressed two-body decays of the leptoquark $U_1$. For the corresponding rates we obtain
\begin{equation}
\begin{aligned}
   \Gamma(U_1\to d_L^i\bar\ell_R^j)
   &= \frac{v^2}{2\Lambda^2}\,\frac{M_{U_1}}{48\pi}\,
    \big| \mathrm{C}^{(0)\,ij}_{\bar Q_L\Phi U_1\ell_R} \big|^2 \,, \\
   \Gamma(U_1\to u_L^i\bar\nu_R^j) 
   &= \frac{v^2}{2\Lambda^2}\,\frac{M_{U_1}}{48\pi}\,
    \big| \mathrm{C}^{(0)\,ij}_{\bar Q_L\tilde\Phi U_1\nu_R} \big|^2 \,, \\
   \Gamma(U_1\to d_R^i\bar\ell_L^j) 
   &= \frac{v^2}{2\Lambda^2}\,\frac{M_{U_1}}{48\pi}\,
    \big| \mathrm{C}^{(0)\,ij}_{\bar d_R\Phi^\dagger U_1 L_L} \big|^2 \,, \\
   \Gamma(U_1\to u_R^i\bar\nu_L^j) 
   &= \frac{v^2}{2\Lambda^2}\,\frac{M_{U_1}}{48\pi}\,
    \big| \mathrm{C}^{(0)\,ij}_{\bar u_R\tilde\Phi^\dagger U_1 L_L} \big|^2 \,, \\
   \Gamma(U_1\to u_L^i\nu_R^j) 
   &= \frac{v^2}{2\Lambda^2}\,\frac{M_{U_1}}{24\pi}\,
    \big| \mathrm{C}^{(0)\,ij}_{\bar Q_L\tilde\Phi U_1\nu_R^c} \big|^2 \,, \\
   \Gamma(U_1\to u_R^i\nu_L^j) 
   &= \frac{v^2}{2\Lambda^2}\,\frac{M_{U_1}}{24\pi}\,
    \big| \mathrm{C}^{(0)\,ij}_{\bar u_R\Phi^\dagger U_1 L_L^c} \big|^2 \,.
\end{aligned}
\end{equation}
Notice that the decay rates from operators containing the field $\slashed{U}_{1v}^\perp$ have a different prefactor compared to decays mediated by operators containing the longitudinal component $\Pi\cdot U_{1v}$, because there are two transverse polarization states of the vector leptoquark. 

For completeness, let us remark at this point that if one wants to extend the construction of the effective Lagrangian for the vector leptoquark $U_1$ to the case of three-jet operators, one needs to generalize the definition \eqref{Pidef} and define two vectors $\Pi_i^\mu$ with $i=1,2$, which satisfy $v\cdot\Pi_i=0$ and $\Pi_i\cdot\Pi_j=-\delta_{ij}$. In the rest frame of the leptoquark, these span the decay plane containing the direction vectors $\vec{n}_1$, $\vec{n}_2$ and $\vec{n}_3$ of the decaying particles.

\section{Scale evolution of the Wilson coefficients}
\label{secwilsoncoeff}

A key strength of any EFT framework is that it allows for a systematic resummation of large logarithmic corrections present in multi-scale problems, which could otherwise spoil the convergence of the perturbative expansion. This is achieved by solving the RG evolution equations of the effective theory. We now discuss the resummation of the large (single and double) logarithms to the two-body decay rates of the leptoquarks $S_1$, $S_3$, and $U_1$, working for simplicity at leading order in SCET power counting. A detailed discussion of the derivation of the anomalous dimensions governing the scale dependence of the Wilson coefficients in the effective Lagrangian, both at leading and subleading order in $\lambda$ (and including the three-jet operators), has been presented in \cite{Alte:2018nbn}.  

At leading power in $\lambda$ there is no operator mixing, and each Wilson coefficient obeys an RG equation of the form
\begin{equation}\label{RGEs}
   \mu\,\frac{d}{d\mu}\,\bm{C}(\mu) = \bm{\Gamma}(\mu)\otimes\bm{C}(\mu) \,,
\end{equation}
where the anomalous-dimension matrix $\bm{\Gamma}$ and each Wilson coefficient $\bm{C}$ are matrices in generation space. The symbol $\otimes$ takes into account that the ordering of $\bm{\Gamma}$ and $\bm{C}$ matters, as will be explained in relation (\ref{matrixeq}) below. For the two-jet operators at $\mathcal{O}(\lambda^2)$ and the three-jet operators at $\mathcal{O}(\lambda^3)$, for which all collinear fields belong to different directions, the all-order expressions for the anomalous dimensions can be derived from a master formula for the anomalous dimensions of scattering amplitudes containing both massless and massive partons derived in \cite{Becher:2009qa,Becher:2009kw}. In color-space notation \cite{Catani:1996vz}, and for the case of relevance to us, in which a single heavy particle $P$ with mass $M_P$ and 4-velocity $v$ decays into several light SM particles with 4-momenta $p_i$, it reads
\begin{equation}\label{master}
   \bm{\Gamma} 
   = \sum_r\Bigg[ \sum_{i<j}\,\bm{T}_i^{(r)}\cdot\bm{T}_j^{(r)} \left( \ln\frac{\mu^2}{m_{ij}^2} + i\pi \right)
    + \sum_i\,\bm{T}_P^{(r)}\cdot\bm{T}_i^{(r)}\,\ln\frac{\mu}{2v\cdot p_i} \Bigg]\,\gamma_{\text{cusp}}^{(r)} 
    + \gamma^P + \sum_i \bm{\gamma}^i \,,
\end{equation}
where the sums are over the final-state particles, and $m_{ij}^2=(p_i+p_j)^2$ is the invariant mass squared of the pair $(ij)$, and $\bm{T}_i^{(r)}$ are the group generators in the representation of particle $i$ (and likewise for the parent particle). The quantity $\gamma_{\text{cusp}}^{(r)}$ is the universal cusp anomalous dimension for light-like Wilson loops in the gauge group $G_r$ \cite{Korchemsky:1987wg,Korchemskaya:1992je,Korchemskaya:1994qp}. Since the SM gauge group is a direct product of three simple groups $G_r$, with $G_1=U(1)_Y$, $G_2=SU(2)_L$, and $G_3=SU(3)_c$, the cusp terms involve a sum over the three groups. Up to next-to-next-to leading order in perturbation theory, $\gamma_{\text{cusp}}^{(r)}$ only depend on the single gauge coupling $\alpha_r$ of the group $G_r$ \cite{Moch:2004pa,Jantzen:2005az}. Note, in particular, that there are no contributions to the cusp anomalous dimensions from Yukawa interactions, because the relevant vertex graphs are found to be power suppressed. Explicitly, one finds
\begin{equation}\label{gammacusp}
\begin{aligned}
   \gamma_{\rm cusp}^{(1)} 
   &= \frac{\alpha_1}{\pi} - \frac{17}{6} \left( \frac{\alpha_1}{\pi} \right)^2 + \dots \,, \\ 
   \gamma_{\rm cusp}^{(2)}
   &= \frac{\alpha_2}{\pi} + \left( 2 - \frac{\pi^2}{6} \right) 
    \left( \frac{\alpha_2}{\pi} \right)^2 + \dots \,, \\ 
   \gamma_{\rm cusp}^{(3)} 
   &= \frac{\alpha_3}{\pi} + \left( \frac{47}{12} - \frac{\pi^2}{4} \right) 
    \left( \frac{\alpha_3}{\pi} \right)^2 + \dots \,,
\end{aligned}   
\end{equation}
where $\alpha_r$ are the three gauge couplings of the SM gauge groups (e.g.\ $\alpha_3=\alpha_s$). In addition to the cusp terms, there are single-particle anomalous dimensions $\gamma^P$ for the leptoquark and $\bm{\gamma}^i$ for the final-state particles, where the latter ones are matrices in generation space. They multiply the corresponding Wilson coefficient either from the left or from the right, as described below. Note that the anomalous dimension in (\ref{anomalousdim}) contains a non-zero imaginary part whenever at least two of the final-state particles are charged under the same gauge group.

It is a simple exercise to specify the general result (\ref{master}) to the case of the leading-power two- or three-jet operators encountered in our analysis.\footnote{The subleading-power two-jet operators contain more than one collinear field moving in the same direction, in which case the form of the anomalous dimensions is more complicated \cite{Alte:2018nbn}.} For the two-jet operators at $\mathcal{O}(\lambda^2)$, we can use charge conservation to obtain the simple formula
\begin{equation}\label{anomalousdim}
   \bm{\Gamma} 
   = - \sum_r \left[ \frac12 \left( C_1^{(r)} + C_2^{(r)} - C_P^{(r)} \right) 
    \left( \ln\frac{\mu^2}{M_P^2} + i\pi \right) + C_P^{(r)}\,\ln\frac{\mu}{M_P} \right] \gamma_{\text{cusp}}^{(r)}
    + \gamma^P + \bm{\gamma}^1 + \bm{\gamma}^2 \,,
\end{equation}
where the symbols $C_P^{(r)}$, $C_1^{(r)}$, $C_2^{(r)}$ denote the eigenvalues of the quadratic Casimir operators of the leptoquark and the $n_1$- and $n_2$-collinear final-state particles, respectively, for the gauge group $G_r$, under which these particles transform. For a non-Abelian group $SU(N)$, the Casimir operator is $C_i=(N^2-1)/2N$ for the fundamental representation and $C_i=N$ for the adjoint representation. For the Abelian group $U(1)_{Y}$, we have instead $C_i=Y_i^2$, where $Y_i$ is the hypercharge of particle $i$.   

Below we discuss the resummation of large logarithmic corrections to the Wilson coefficients in the mass basis for the leading-power two-jet operators for the leptoquarks $S_1$, $S_3$, and $U_1$. We work at leading order in RG-improved perturbation theory, which is equivalent to resumming the large logarithms at next-to-leading logarithmic order. This requires the two-loop expressions for cusp anomalous dimensions and the SM $\beta$-functions along with one-loop expression for $\gamma^P$ and $\bm{\gamma}^i$. This allows us to consistently estimate the leading resummation effects to the various tree-level decay rates. It is not difficult to extend our formalism to the case of the leading-power three-jet operators. For the case of the leptoquark $S_1$, the corresponding anomalous dimensions are collected in Appendix~\ref{anomalousdim3jet}.

\subsection[Resummation effects for decays of the scalar leptoquark $S_1$]{\boldmath Resummation effects for decays of the scalar leptoquark $S_1$}  
\label{resumationS1}

According to the master formula (\ref{anomalousdim}), the anomalous dimensions of the two-jet operators in (\ref{S1operators}) are
\begin{equation}\label{GammaS1LP}
\begin{aligned}
   \bm{\Gamma}_{\bar u_R^c\ell_R S_1^\ast} 
   &= - \frac{2}{3} \gamma^{(1)}_{\text{cusp}} \left( \ln\frac{\mu^2}{M^2_{S_1}} + i\pi \right) 
    - \left( \frac{4}{3} \gamma^{(3)}_{\text{cusp}} + \frac{1}{9} \gamma^{(1)}_{\text{cusp}} \right)
    \ln\frac{\mu}{M_{S_1}} \\
   &\quad + \gamma^{S_1} +\left( \mathbb{\gamma}^{\ell_R},\,.\,\right) + \left(.\,,\bm{\gamma}^{u_R}\right) , \\
   \bm{\Gamma}_{\bar Q^c_L L_L S_1^\ast} 
   &= - \left( \frac{3}{4} \gamma^{(2)}_{\text{cusp}} + \frac{1}{12} \gamma^{(1)}_{\text{cusp}} \right)
    \left( \ln\frac{\mu^2}{M^2_{S_1}} + i\pi \right)
    - \left( \frac{4}{3} \gamma^{(3)}_{\text{cusp}} + \frac{1}{9} \gamma^{(1)}_{\text{cusp}} \right) 
    \ln\frac{\mu}{M_{S_1}} \\
   &\quad + \gamma^{S_1} + \left(\bm{\gamma}^{L_L},\,.\right) + \left(.\,,\bm{\gamma}^{Q_L}\right) ,\\
   \bm{\Gamma}_{\bar d_R\nu_R S_1^\ast} 
   &= - \left( \frac{4}{3} \gamma^{(3)}_{\text{cusp}} + \frac{1}{9} \gamma^{(1)}_{\text{cusp}} \right) 
    \ln\frac{\mu}{M_{S_1}} + \gamma^{S_1} + \left(.\,,\bm{\gamma}^{d_R}\right) ,
\end{aligned}
\end{equation} 
where we use the notations $\left( .\,,\bm{\gamma}\right)$ and $\left(  \bm{\gamma}\,,.\right)$ for the single-particle anomalous dimensions to indicate a multiplication with the Wilson coefficient from the left and from the right, respectively, such that
\begin{equation}\label{matrixeq}
\begin{aligned}
   \left(.\,,\bm{\gamma}\right)\otimes\bm{C}
   &\equiv \bm{C}\,\bm{\gamma} \,, \\
   \left(\bm{\gamma}\,,.\right)\otimes\bm{C}
   &\equiv \bm{\gamma}\,\bm{C} \,.
\end{aligned}
\end{equation} 
The various single-particle anomalous dimensions in (\ref{GammaS1LP}) are \cite{Alte:2018nbn}
\begin{align}\label{fieldanomdim}
   \bm{\gamma}^{\ell_R}
   &= - \frac{\alpha_1}{4\pi} + \frac{1}{16\pi^2}\,\bm{Y}_\ell^\dagger\bm{Y}_\ell \,, \notag\\
   \bm{\gamma}^{L_L}
   &= - \frac{9\alpha_2}{16\pi} - \frac{\alpha_1}{16\pi} + \frac{1}{32\pi^2}\,\bm{Y}_\ell\bm{Y}_\ell^\dagger \,,
    \notag\\
   \bm{\gamma}^{u_R}
   &= - \frac{\alpha_3}{\pi} - \frac{\alpha_1}{9\pi} + \frac{1}{16\pi^2}\,\bm{Y}_u^\dagger\bm{Y}_u \,, \\
   \bm{\gamma}^{d_R}
   &= - \frac{\alpha_3}{\pi} - \frac{\alpha_1}{36\pi} + \frac{1}{16\pi^2}\,\bm{Y}_d^\dagger\bm{Y}_d \,, \notag\\
   \bm{\gamma}^{Q_L}
   &= - \frac{\alpha_3}{\pi} - \frac{9\alpha_2}{16\pi} - \frac{\alpha_1}{144\pi}
    + \frac{1}{32\pi^2} \left( \bm{Y}_u\bm{Y}_u^\dagger + \bm{Y}_d\bm{Y}_d^\dagger \right) , \notag
\end{align}
where $\bm{Y}_\ell$ is the Yukawa matrix for the leptons, while $\bm{Y}_u$ and $\bm{Y}_d$ are the Yukawa matrices for the up- and down-type quarks. The anomalous dimension of the heavy scalar $S_1$ is \cite{Becher:2009kw}
\begin{equation}
   \gamma^{S_1} = - \frac{2\alpha_3}{3\pi} - \frac{\alpha_1}{18\pi} \,.
\end{equation}
In practice, we transform the Wilson coefficients to the mass basis, since this is the relevant basis for physical quantities such as decay rates. In the mass basis the Yukawa matrices in (\ref{fieldanomdim}) become diagonal except for the case of $\bm{\gamma}^{Q_L}$. In this case one needs to distinguish between the up-type quarks and the down-type quarks in the doublet \cite{Heiles:2020plj}, for which
\begin{equation}\label{gammaleft}
\begin{aligned}
   \bm{\gamma}^{u_L}
   &= - \frac{\alpha_3}{\pi} - \frac{9\alpha_2}{16\pi} - \frac{\alpha_1}{144\pi}
    + \frac{1}{32\pi^2}\,\Big[ \text{diag}\left( y_u^2, y_c^2, y_t^2 \right)
    + \bm{V} \text{diag}\left( y_d^2, y_s^2, y_b^2 \right)\!\bm{V}^\dagger \Big] \,, \\
   \bm{\gamma}^{d_L}
   &= - \frac{\alpha_3}{\pi} - \frac{9\alpha_2}{16\pi} - \frac{\alpha_1}{144\pi}
    + \frac{1}{32\pi^2}\,\Big[ \bm{V}^\dagger \text{diag}\left( y_u^2, y_c^2, y_t^2 \right) \bm{V}
    + \text{diag}\left( y_d^2, y_s^2, y_b^2 \right) \Big] \,,
\end{aligned}
\end{equation}  
where $y_q$ denotes the Yukawa coupling of the quark mass eigenstate q, and $\bm{V}$ is the Cabibbo--Kobayashi--Maskawa  matrix. For the numerical estimates we take into account only the top-quark Yukawa coupling. All the other quark Yukawa couplings would have a tiny impact on the resummation effects. We also neglect the Yukawa couplings of the leptons. The evolution of the Yukawa coupling of the top quark is given by \cite{Grzadkowski:1987tf}
\begin{equation}
   \mu\,\frac{d}{d\mu}\,y_t(\mu) 
   = \frac{9 y_t^3}{32\pi^2} 
    - y_t \left( \frac{17\alpha_1}{48\pi} + \frac{9\alpha_2}{16\pi} + \frac{2\alpha_3}{\pi} \right) .
\end{equation}

\begin{figure}[t]
\begin{center}
\includegraphics[width=0.69\textwidth]{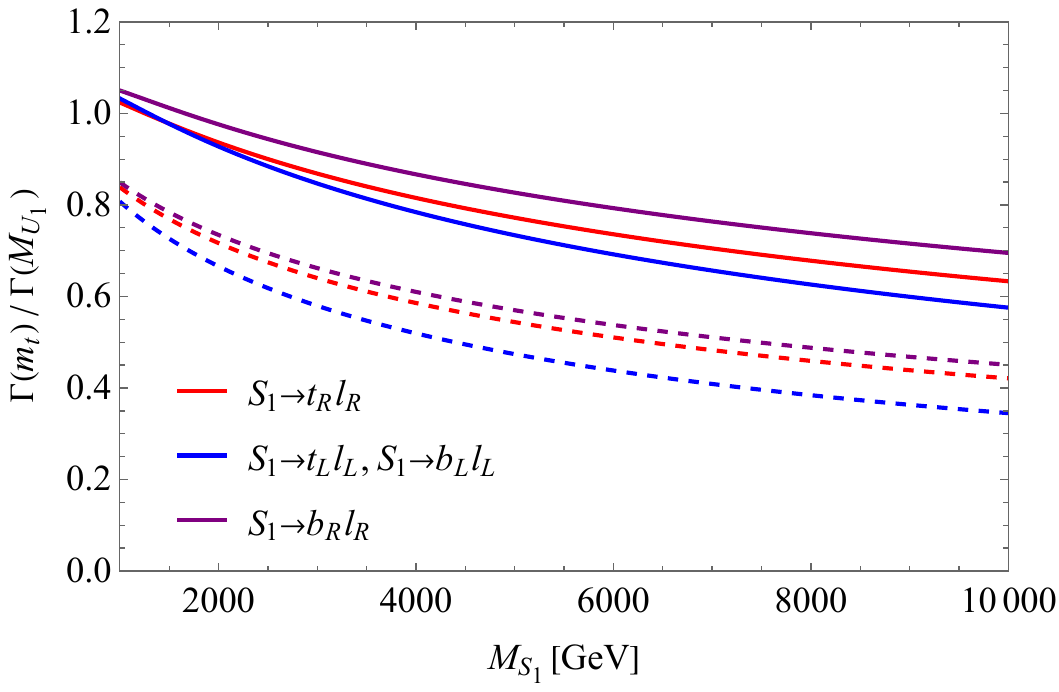}
\caption{\label{fig:graph1}
Resummation effects on the tree-level decay rates for the two-jet decays $S_1\to t_R\ell_R$ (red), $S_1\to t_L\ell_L$ and $S_1\to b_L\ell_L$ (blue), and $S_1\to b_R\ell_R$ (purple) as a function of the leptoquark mass $M_{S_1}$. The relevant Wilson coefficients are evolved from the scale of the leptoquark mass down to $\mu=m_t$. The solid lines show the results of a complete numerical solution of the RG equations, while the dashed curves are obtained by retaining only the cusp terms in the anomalous dimensions.}
\end{center}
\end{figure}
 
We now present numerical results for the resummation effects on the Wilson coefficients of the leading-power two-jet operators for the scalar leptoquark $S_1$ in (\ref{S1operators}). For these operators the largest effects arise for decays into third-generation quarks, for which the top-quark Yukawa coupling plays an important role. We fix the low scale to the top quark mass and consider a leptoquark with mass $M_{S_1}=3$~TeV. From a numerical integration of the RG evolution equation (\ref{RGEs}) for each coefficient, we obtain (for each lepton flavor $j=1,2,3$) 
\begin{equation}
\begin{aligned}
   \mathrm{C}_{\bar u_R^c\ell_R S_1^\ast}^{3j}(m_t)
   &\approx 0.93\,e^{0.02 i}\,\mathrm{C}_{\bar u_R^c\ell_R S_1^\ast}^{3j}(M_{S_1}) \,, \\
   \mathrm{C}_{\bar Q_L^c L_L S_1^\ast}^{3j}(m_t)
   &\approx 0.92\,e^{0.07 i}\,\mathrm{C}_{\bar Q_L^c L_L S_1^\ast}^{3j}(M_{S_1}) \,, \\
   \mathrm{C}_{\bar d_R\nu_R S_1^\ast}^{3j}(m_t)
   &\approx 0.96\,\mathrm{C}_{\bar d_R\nu_R S_1^\ast}^{3j}(M_{S_1}) \,.
\end{aligned}
\end{equation}
The corresponding rates for the decays $S_1\to t_R\ell_R$, $S_1\to t_L\ell_L$ and $S_1\to b_L\ell_L$, and $S_1\to b_R\ell_R$ are proportional to the absolute squares of the Wilson coefficients. In Figure~\ref{fig:graph1}, we show the resummation effects on these rates as a function of $M_{S_1}$. The solid lines show the reduction factors obtained from the full solution of the evolution equations, while the dashed lines show for comparison the double-logarithmic contributions from ``cusp terms'', neglecting the single-logarithmic effects from the single-particle anomalous dimensions. In practice, this approximation is often inherent in parton showers. Importantly, however, it can be seen from Figure~\ref{fig:graph1} that neglecting the single-logarithmic terms gives a poor numerical approximation to the full results. In fact, it is an important merit of our EFT approach that RG methods allow for a consistent resummation of all large logarithmic corrections.   

\subsection[Resummation effects for decays of the scalar leptoquark $S_3$]{\boldmath Resummation effects for decays of the scalar leptoquark $S_3$}

As shown in (\ref{S3LPdecays}), at leading power there are four possible two-jet decay channels for the leptoquark $S_3$, whose rates are governed by a single Wilson coefficient $C_{\bar Q^c_L S_3^\ast L_L}^{ij}$. Its anomalous dimension is obtained as
\begin{equation}
\begin{aligned}
   \bm{\Gamma}_{\bar Q_L^c S_3^\ast L_L}
   &= \left( \frac{1}{4} \gamma_{\text{cusp}}^{(2)} - \frac{1}{12} \gamma_{\text{cusp}}^{(1)} \right)
    \left( \ln\frac{\mu^2}{M_{S_3}^2} + i\pi \right)
   - \left( \frac{4}{3} \gamma^{(3)}_{\text{cusp}} + 2 \gamma_{\text{cusp}}^{(2)} 
    + \frac{1}{9} \gamma_{\text{cusp}}^{(1)} \right) \ln\frac{\mu}{M_{S_3}} \\
   &\quad + \gamma^{S_3} + \left(\bm{\gamma}^{L},.\right) + \left(.\,,\bm{\gamma}^{Q}\right) ,
\end{aligned}
\end{equation} 
where $\bm{\gamma}^{L_L}$ and $\bm{\gamma}^{Q_L}$ have been given in (\ref{fieldanomdim}), and \cite{Becher:2009kw}
\begin{equation}
   \gamma^{S_3} = - \frac{2\alpha_3}{3\pi} - \frac{\alpha_2}{\pi} - \frac{\alpha_1}{18\pi} \,.
\end{equation}

\begin{figure}[t]
\begin{center}
\includegraphics[width=0.69\textwidth]{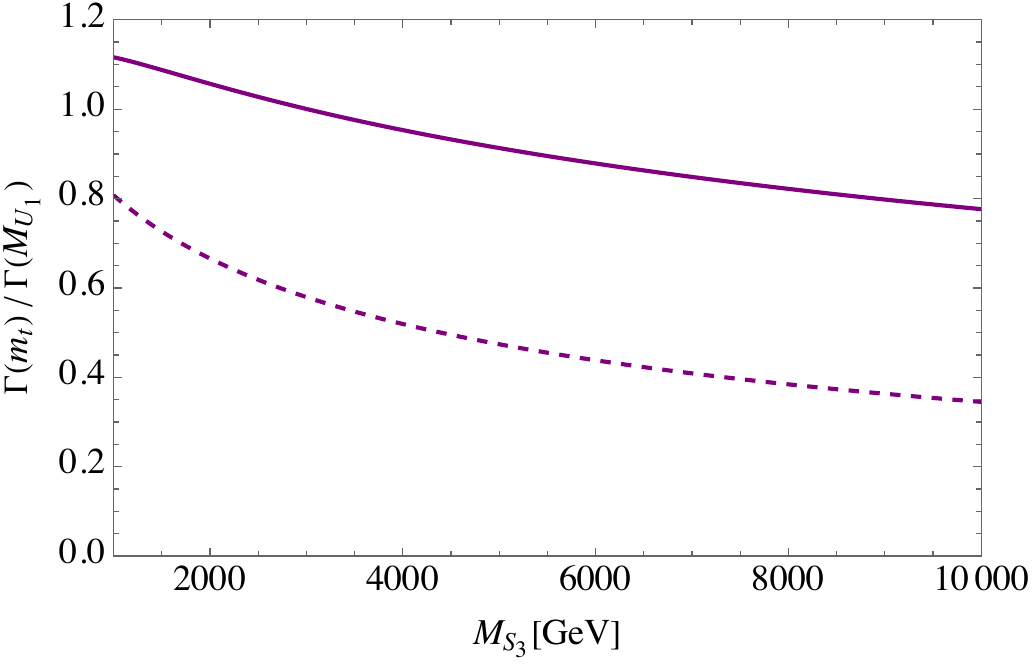}
\end{center}
\caption{\label{fig:graph2}
Resummation effects on the tree-level decay rates for the two-jet decays $S_3^{2/3}\to t_L\nu_L$, $S_3^{-4/3}\to b_L\ell_L$, $S_3^{-1/3}\to b_L\nu_L$ and $S_3^{-1/3}\to t_L\ell_L$, all of which are governed by the same Wilson coefficient, as a function of the leptoquark mass $M_{S_3}$. The Wilson coefficient is evolved from the scale of the leptoquark mass down to $\mu=m_t$. The solid line shows the results of a complete numerical solution of the RG equation, while the dashed curve is obtained by retaining only the cusp terms in the anomalous dimension.}
\end{figure}

For a 3~TeV leptoquark the effects of scale evolution are smaller in this case than for the case of $S_1$. They become more seizable for larger leptoquark masses. For instance, with $M_{S_3}=4.5$~TeV we find
\begin{equation}
   \mathrm{C}_{\bar Q_L^c S_3^\ast L_L}^{3j}(m_t)
   \approx 0.97\,e^{-0.02 i}\,\mathrm{C}_{\bar Q_L^c S_3^\ast L_L}^{3j}(M_{S_3}) \,.
\end{equation} 
In Figure~\ref{fig:graph2}, we show the mass dependence of the resummation effects for the four decay rates governed by this Wilson coefficient. Comparing the solid line with the dashed one, we see that it would again be a poor approximation to only retain the cusp terms in the anomalous dimension.

\subsection[Resummation effects for decays of the vector leptoquark $U_1$]{\boldmath Resummation effects for decays of the vector leptoquark $U_1$}

In a similar fashion, we can derive the anomalous dimensions of the leading-order two-jet operators for the leptoquark $U_1$ shown in (\ref{leadingU1operators}). We find
\begin{align}
   \bm{\Gamma}_{\bar Q_L U_1 L_L}
   &= \left( - \frac{3}{4} \gamma_{\text{cusp}}^{(2)} + \frac{1}{12} \gamma^{(1)}_{\text{cusp}}\right)
    \left( \ln\frac{\mu^2}{M_{U_1}^2} + i\pi \right)
    - \left( \frac{4}{3} \gamma^{(3)}_{\text{cusp}} + \frac{4}{9} \gamma^{(1)}_{\text{cusp}} \right)
    \ln\frac{\mu}{M_{U_1}} \notag\\
   &\quad + \gamma^{U_1} + \left(\bm{\gamma}^{L}\,,.\right) + \left(.\,,\bm{\gamma}^{Q}\right) , \notag\\
   \bm{\Gamma}_{\bar d_R U_1\ell_R}
   &= - \frac{1}{3} \gamma^{(1)}_{\text{cusp}} \left( \ln\frac{\mu^2}{M_{U_1}^2} + i\pi \right)
    - \left( \frac{4}{3} \gamma^{(3)}_{\text{cusp}} + \frac{4}{9} \gamma^{(1)}_{\text{cusp}} \right) 
    \ln\frac{\mu}{M_{U_1}} 
    + \gamma^{U_1} + \left(\bm{\gamma}^{\ell_R}\,,.\right) + \left(.\,,\bm{\gamma}^{d_R}\right) , \notag\\
   \bm{\Gamma}_{\bar u_R U_1\nu_R}
   &= - \left( \frac{4}{3} \gamma^{(3)}_{\text{cusp}} + \frac{4}{9} \gamma_{\text{cusp}}^{(1)} \right)
    \ln\frac{\mu}{M_{U_1}} +\gamma^{U_1} + \left(.\,,\bm{\gamma}^{u_R}\right) ,
\end{align}
where the anomalous dimension of the leptoquark $U_1$ reads \cite{Becher:2009kw}
\begin{equation}
   \gamma^{U_1} = - \frac{2\alpha_3}{3\pi} - \frac{2\alpha_1}{9\pi} \,.
\end{equation}

\begin{figure}[t]
\begin{center}
\includegraphics[width=0.69\textwidth]{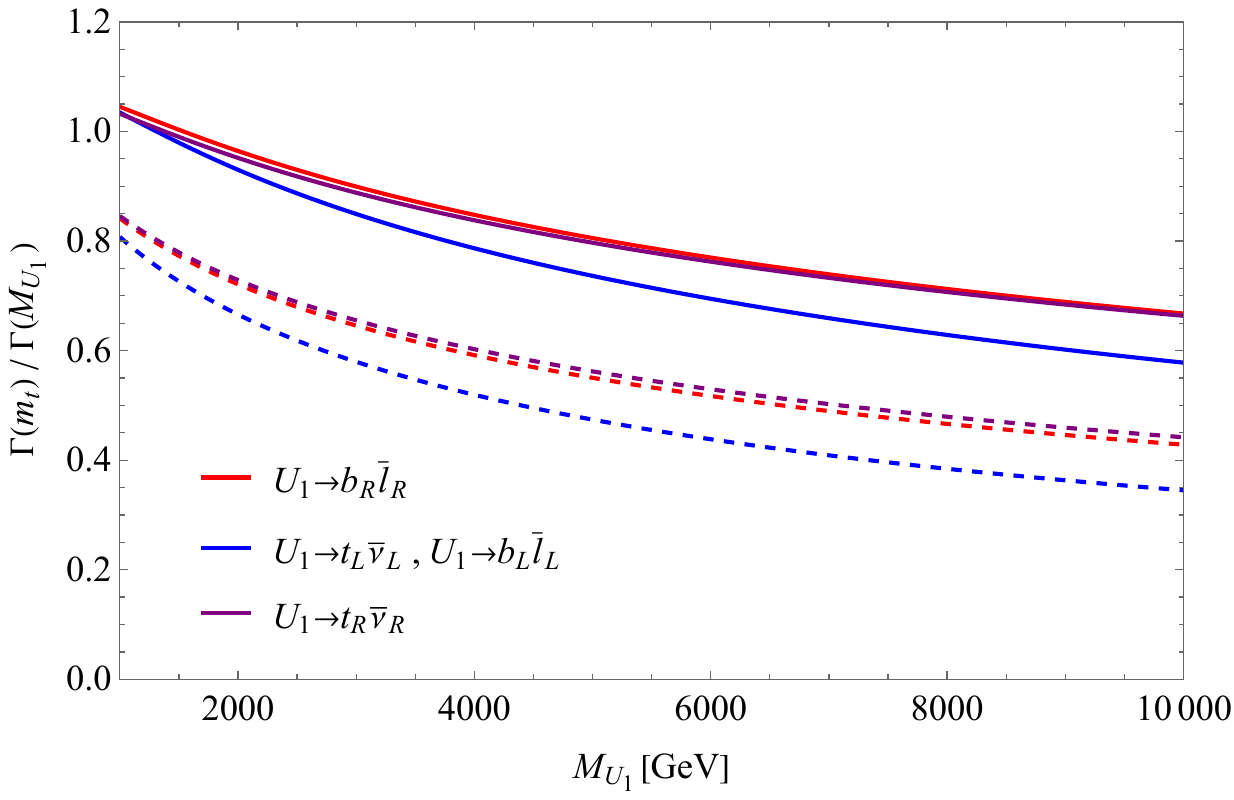}
\end{center}
\caption{\label{fig:graph3}
Resummation effects on the tree-level decay rates for the two-jet decays $U_1\to b_R\bar\ell_R$ (red), $U_1\to t_L\bar\nu_L$ and $U_1\to b_L\bar\ell_L$ (blue), and $U_1\to t_R\bar\nu_R$ (purple) as a function of the leptoquark mass $M_{U_1}$. The relevant Wilson coefficient are evolved from the scale of the leptoquark mass down to $\mu=m_t$. The solid lines show the results of a complete numerical solution of the RG equations, while the dashed curves are obtained by retaining only the cusp terms in the anomalous dimensions.}
\end{figure}

For decays into third-generation quarks and a leptoquark mass $M_{U_{1}}=3$~TeV, we find for the relevant Wilson coefficients the evolution effects
\begin{equation}
\begin{aligned}
   \mathrm{C}_{\bar Q_L U_1 L_L}^{3j}(m_t)
   &\approx 0.92\,e^{0.06 i}\,\mathrm{C}_{\bar Q_L U_1 L_L}^{3j}(M_{U_1}) \,, \\
   \mathrm{C}_{\bar d_R U_1\ell_R}^{3j}(m_t)
   &\approx 0.95\,e^{0.01 i}\,\mathrm{C}_{\bar d_R U_1\ell_R}^{3j}(M_{U_1}) \,, \\
   \mathrm{C}_{\bar u_R U_1\nu_R}^{3j}(m_t)
   &\approx 0.94\,\mathrm{C}_{\bar u_R U_1\nu_R}^{3j}(M_{U_1}) \,.
\end{aligned}
\end{equation}
The coefficients govern the two-jet decay rates $U_1\to t_L\bar\nu_L$ and $U_1\to b_L\bar\ell_L$, $U_1\to b_R\bar\ell_R$, and $U_1\to t_R\bar\nu_R$, respectively. The resummation effects for the corresponding decay rates are depicted in Figure~\ref{fig:graph3}. Also in this case there is a significant difference between the full results and those obtained by neglecting the single-particle anomalous dimensions.

\section{Tree-level matching conditions in specific models}
\label{sectionmatching}

In this section we briefly look at some concrete UV models for each of the three leptoquarks we have considered and match them at tree level to the corresponding SCET Lagrangians. We discuss the matching procedure for the two-jet operators at leading and subleading order in the expansion parameter $\lambda$. 

\subsection[Matching conditions for the scalar leptoquark $S_1$]{\boldmath Matching conditions for the scalar leptoquark $S_1$}

We start with the renormalizable Lagrangian 
\begin{equation}\label{S1UV}
\begin{aligned}
   \mathcal{L}_{S_1}
   &= (D_\mu S_1)^\dagger (D^\mu S_1) - M_{S_1}^2 S_1^\dagger S_1 \\
   &\quad + \Big[ g_{1L}^{ij}\,\bar Q_L^{c,i}\,i\sigma_2 L_L^j S_1^\star
    + g_{1R}^{ij}\,\bar u_R^{c,i} \ell_R^j S_1^\star 
    + g_{1\nu}^{ij}\,\bar d_R^{c,i}\nu_R^j S_1^\star + \text{h.c.} \Big] \,,
\end{aligned}
\end{equation}
where compared with e.g.\ \cite{Sakaki:2013bfa,Bauer:2015knc} we have included an additional term for a right-handed neutrino. These dimension-4 interaction terms have the same structure as the operators in our leading-order effective Lagrangian in \eqref{Lagrangian1}. The tree-level matching conditions at the high scale $\mu_h\approx M_{S_1}$ are therefore trivially given as
\begin{equation}
\begin{aligned}
   C_{\bar u_R^c\ell_R S_1^\ast}^{ij}(\Lambda,M_{S_1},\mu_h) &= g_{1R}^{ij} \,, \\
   C_{\bar Q_L^c L_L S_1^\ast}^{ij}(\Lambda,M_{S_1},\mu_h) &= g_{1L}^{ij} \,, \\
   C_{\bar d_R^c\nu_R S_1^{\ast}}^{ij}(\Lambda,M_{S_1},\mu_h) &= g_{1\nu}^{ij} \,.
\end{aligned}
\end{equation}
Beyond tree-level, the Wilson coefficients receive non-trivial matching conditions from hard modes at the scale of the leptoquark mass, which are integrated out in the construction of the low-energy EFT.  

Since all interactions in \eqref{S1UV} change fermion number by two units, while the two-jet operators appearing at subleading power in $\lambda$ conserve fermion number, see \eqref{Eq.1} and \eqref{Eq.2}, we trivially obtain the tree-level matching conditions
\begin{equation}
\begin{aligned}
   C_{\bar d_R\tilde\Phi^\dagger L_L S_1}^{(0)\,ij}(\Lambda,M_{S_1},\mu_h)
   &= C_{\bar d_R\tilde\Phi^\dagger L_L S_1}^{(k)\,ij}(\Lambda,M_{S_1},\mu_h) = 0 \,, \\
   C_{\bar Q_L\Phi\nu_R S_1}^{(0)\,ij}(\Lambda,M_{S_1},\mu_h)
   &= C_{\bar Q_L\Phi\nu_R S_1}^{(k)\,ij}(\Lambda,M_{S_1},\mu_h) = 0 \,, \\
   C_{\bar d_R B\nu_R S_1}^{(k)\,ij}(\Lambda,M_{S_1},\mu_h)
   &= 0 \,,
\end{aligned}
\end{equation}
with $k=1,2$.

\begin{figure}[t]
\center
\includegraphics[scale=0.4]{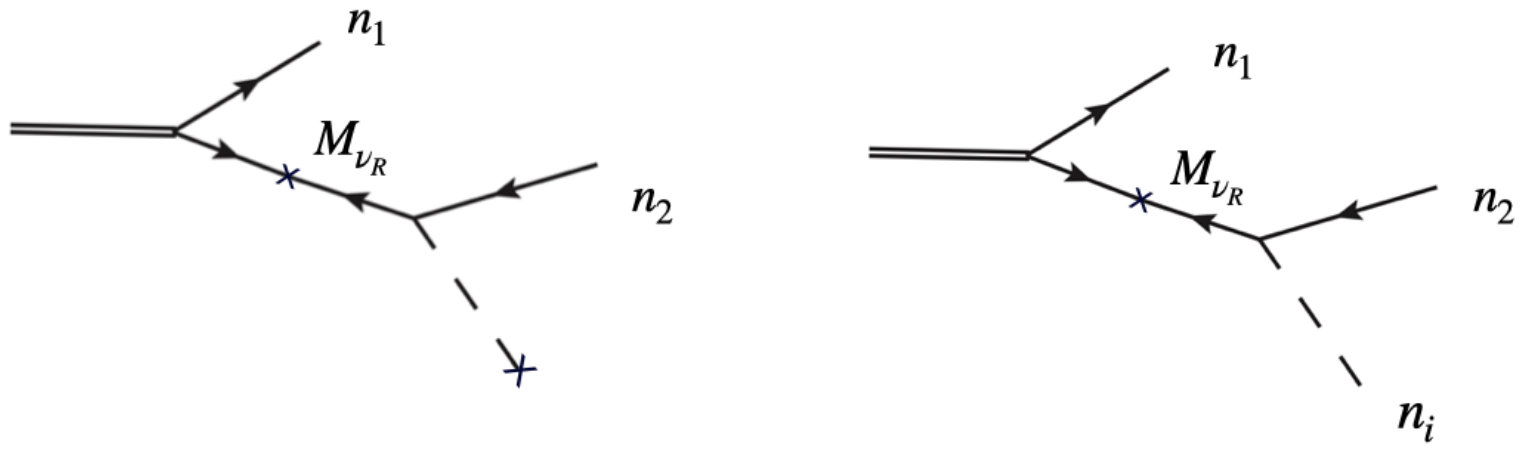} 
\caption{\label{fig:DeltaF=2} 
Tree-level Feynman diagrams involving a Majorana mass insertions on the neutrino propagator in models with a heavy right-handed neutrino. A double line represents the leptoquark, a dashed line with a cross the Higgs vacuum expectation value, and a dashed line a Higgs boson. The final-state particle moving along $n_1$ is a quark, while the particle moving along $n_2$ is a lepton. The Higgs bosons can move in either direction.}
\end{figure}

This last conclusion can be altered if the UV theory contains additional new interactions beyond those mediated by the leptoquark $S_1$, and if these interactions violate fermion number. As a prototypical example, we consider a model featuring a {\em heavy\/} right-handed neutrino with mass $M_\nu\gtrsim M_{S_1}$, which is no longer a light degree of freedom, but which can mediate fermion-number violating interactions in the UV. Then the diagrams shown in Figure~\ref{fig:DeltaF=2} give rise to non-zero matching coefficients for the operators $\mathcal{O}_{\bar d_R\tilde\Phi^\dagger L_L S_1}^{(n)\,ij}$, for which we find
\begin{equation}
\begin{aligned}
   C_{\bar d_R\tilde\Phi^\dagger L_L S_1}^{(0)\,ij}(\Lambda,M_{S_1},\mu_h)
   &= C_{\bar d_R\tilde\Phi^\dagger L_L S_1}^{(2)\,ij}(\Lambda,M_{S_1},\mu_h,u) 
    = - \frac{1}{M_\nu} \left( \bm{g}_{1\nu}^\ast\bm{Y}_\nu^\dagger \right)^{ij} , \\
   C_{\bar d_R\tilde\Phi^\dagger L_L S_1}^{(1)\,ij}(\Lambda,M_{S_1},\mu_h,u)
   &= - \frac{M_\nu}{M_\nu^2-u M_{S_1}^2} \left( \bm{g}_{1\nu}^\ast\bm{Y}_\nu^\dagger \right)^{ij} .
\end{aligned}
\end{equation}
This shows that it is important and in some cases even crucial to include the $\mathcal{O}(\lambda^3)$ two-jet operators in the effective Lagrangian. For instance, if the couplings $g_{1L}^{ij}$ and $g_{1R}^{ij}$ in the Lagrangian \eqref{S1UV} vanish for some reason and only the coefficients $g_{1\nu}^{ij}$ are non-zero, then at leading power only the decay $S_1\to d_R^i\nu_R^j$ is allowed, but if $M_\nu>M_{S_1}$ this channel is kinematically not accessible. At subleading power, however, the presence of the coefficient $C_{\bar d_R\tilde\Phi^\dagger L_L S_1}^{(0)\,ij}$ allows the decay mode $S_1\to d_R^i\bar\nu_L^j$, see \eqref{S12bodysuppressed}, which would manifest itself as a mono-jet signature with large missing energy. 

\subsection[Matching conditions for the scalar leptoquark $S_3$]{\boldmath Matching conditions for the scalar leptoquark $S_3$}

The renormalizable interactions of the scalar leptoquark $S_3$ with SM fields are given by
\begin{equation}\label{S3UVL}
   \mathcal{L}_{S_3}
   = (D_\mu S_3^a)^\dagger (D^\mu S_3^a) - M_{S_3}^2 S_3^{a\dagger} S_3^a
    + \big[ g_{3L}^{ij}\,\bar Q_L^{c,i}\,i\sigma_2\,S_3^\ast L_L^j + \text{h.c.} \big] \,,
\end{equation}
which once again is of the same form as the leading-power two-jet operator in \eqref{LOopS3}. At a high scale $\mu_h\approx M_{S_3}$, we thus obtain the tree-level matching condition
\begin{equation}
   C_{\bar Q_L^c S_3^\ast L_L}^{ij}(\Lambda,M_{S_1},\mu_h) = g_{3L}^{ij} \,.
\end{equation}
Also in this case the Wilson coefficients of all two-jet operators that appear at subleading power in (\ref{S32jetsub}) vanish, because the corresponding operators conserve fermion number. We thus have
\begin{equation}
\begin{aligned}
   C_{\bar Q_L S_3\Phi\nu_R}^{(0)\,ij}(\Lambda,M_{S_1},\mu_h)
   &= C_{\bar Q_L S_3\Phi\nu_R}^{(k)\,ij}(\Lambda,M_{S_1},\mu_h) = 0 \,, \\
   C_{\bar d_R\tilde\Phi^\dagger S_3 L_L}^{(0)\,ij}(\Lambda,M_{S_1},\mu_h)
   &= C_{\bar d_R\tilde\Phi^\dagger S_3 L_L}^{(k)\,ij}(\Lambda,M_{S_1},\mu_h) = 0 \,, \\
   C_{\bar d_R S_3 W\nu_R}^{(k)\,ij}(\Lambda,M_{S_1},\mu_h)
   &= 0 \,,
\end{aligned}
\end{equation}
with $k=1,2$. In this case, extending the model with a heavy-right-handed neutrino does not give rise to non-zero matching coefficients of the $\mathcal{O}(\lambda^3)$ two-jet operators.

\subsection[Matching conditions for the vector leptoquark $U_1$]{\boldmath Matching conditions for the vector leptoquark $U_1$}

We finally explore a UV completion of the SM that describes the interactions of the vector leptoquark $U_1$. Since this requires a massive vector field, the model must contain a dynamical mechanism for giving mass to the leptoquark, which requires adding more fields in addition to the field $U_1$ itself, such as additional heavy scalars and (vector-like) fermions. The cases usually considered in the literature are either ``gauge models'' or ``strongly interacting models'' \cite{Pati:1974yy,DiLuzio:2017vat,DiLuzio:2018zxy}. In gauge models, the $U_1$ leptoquark is the massive gauge boson of a spontaneously broken gauge symmetry $G_{\rm NP}\ni G_{\rm SM}$. In strongly interacting models, on the other hand, the $U_1$ appears as a massive resonance for a new, strongly interacting sector. Here we will be agnostic to the presence of additional new fields beyond the leptoquark $U_1$ and assume that the corresponding particles are heavier than the leptoquark and can be integrated out. The most general Lagrangian describing the interaction of only the leptoquark $U_1$ with the SM reads \cite{DiLuzio:2018zxy,Baker:2019sli} 
\begin{equation}\label{LU1}
\begin{aligned}
   \mathcal{L}_{U_1} 
   &= - \frac{1}{2} \left( D^\mu U_1^\nu - D^\nu U_1^\mu \right)^\dagger
    \left( D_\mu U_{1\nu} - D_\nu U_{1\mu} \right) + M_{U_1}^2 U_{1\mu}^\dagger U_1^\mu \\
   &\quad + \left[ \frac{g_U}{\sqrt2} \left( \beta_L^{ij}\,\bar Q_L^i\slashed{U}_{\!1} L_L^j 
    + \beta_R^{ij}\,\bar d_R^i\slashed{U}_{\!1}\ell_R^j \right) 
    + g_U^\nu\,\beta_\nu^{ij}\,\bar u_R^i\slashed{U}_{\!1}\nu_R^j + \text{h.c.} \right] \\
   &\quad - i g_3 \left( 1-\kappa_U \right) U_{1\mu}^\dagger G^{\mu\nu} U_{1\nu} 
    - i\,\mathcal{Y}_{U_1}\,g_1 \left( 1-\tilde\kappa_U \right) U_{1\mu}^\dagger B^{\mu\nu} U_{1\nu} \,,
\end{aligned}
\end{equation}
where $\mathcal{Y}_{U_1}=\frac23$ is the hypercharge of the leptoquark, and compared with \cite{Baker:2019sli} we have added an interaction term involving the right-handed neutrino. The gauge-invariant terms shown in the last line are present in gauge models, for which $\kappa_U=\tilde\kappa_U=0$. In strongly interacting models, and assuming the so-called minimal-coupling scenario, these terms vanish, i.e., $\kappa_U=\tilde\kappa_U=1$.

\begin{figure}[t]
\center
\includegraphics[scale=0.5]{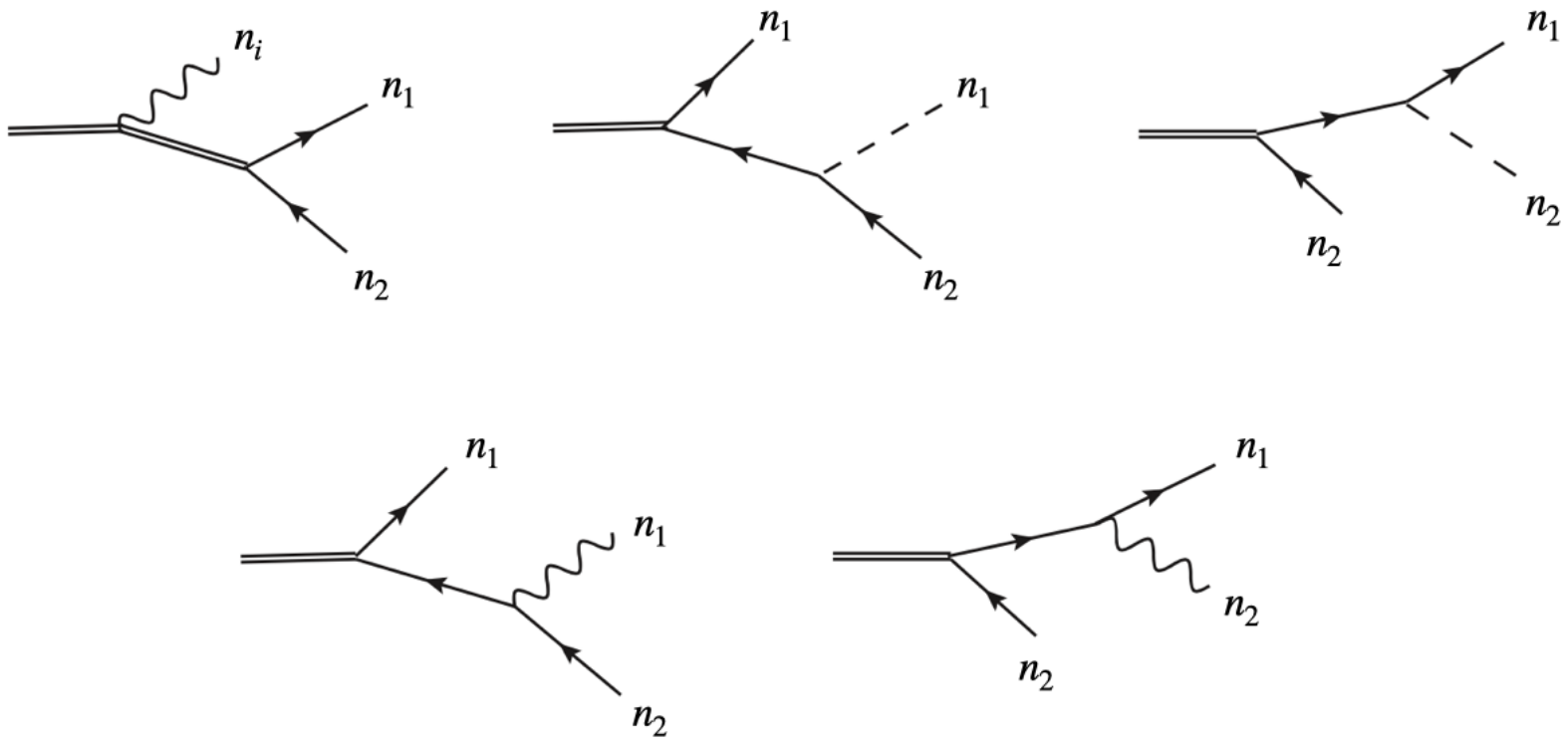} 
\caption{\label{matchingLQdiag} 
Tree-level Feynman diagrams needed for the calculation of the matching conditions for the Wilson coefficients of the subleading-power two-jet operators for the leptoquark $U_1$. A double line represents the leptoquark, a dashed line a Higgs boson, and a wavy line a SM gauge boson. The final-state particle moving along $n_1$ is a quark, while the particle moving along $n_2$ is a lepton. The gauge boson in the first diagram can move along either direction.}
\end{figure}

The tree-level matching conditions of the leading-power two-jet operators in \eqref{LagrangianU1} at a matching scale $\mu_h\approx M_{U_1}$ are readily obtain as
\begin{equation}
\begin{aligned}
   C_{\bar Q_L U_1 L_L}^{ij}(\Lambda,M_{U_1},\mu_h) 
   &= \frac{g_U}{\sqrt2}\,\beta_L^{ij} \,, \\
   C_{\bar d_R U_1\ell_R}^{ij}(\Lambda,M_{U_1},\mu_h) 
   &= \frac{g_U}{\sqrt2}\,\beta_R^{ij} \,, \\
   C_{\bar u_R U_1\nu_R}^{ij}(\Lambda,M_{U_1},\mu_h) 
   &= g_U^\nu\beta_\nu^{ij} \,.
\end{aligned}
\end{equation}

In the case of the leptoquark $U_1$, we find that most of the Wilson coefficients of the two-jet operators appearing at subleading power are also non-zero. The relevant Feynman diagrams needed to calculate the corresponding tree-level matching conditions are shown in Figure~\ref{matchingLQdiag}. Note that the emitted final-state bosons must carry a large momentum in the direction opposite to the particles to which they are attached, since only in this case one gets a hard propagator, which can be integrated out. Focussing first on the operators containing gauge fields, defined in \eqref{U1subLa}, we obtain
\begin{align}
   \frac{1}{\Lambda}\,C_{\bar Q_L G U_1 L_L}^{(1)\,ij}(\Lambda,M_{U_1},\mu_h,u) 
   &= - \left( 2 - \kappa_U \right) \frac{g_U}{\sqrt2 M_{U_1}}\,\beta_L^{ij} \,, \notag\\
   \frac{1}{\Lambda}\,C_{\bar Q_L W U_1 L_L}^{(1)\,ij}(\Lambda,M_{U_1},\mu_h,u) 
   &= \frac{g_U}{\sqrt2 M_{U_1}}\,\beta_L^{ij} \,, \notag\\
   \frac{1}{\Lambda}\,C_{\bar Q_L B U_1 L_L}^{(1)\,ij}(\Lambda,M_{U_1},\mu_h,u)  
   &= \left[ \mathcal{Y}_{L_L} - \left( 2 - \tilde\kappa_U \right) \mathcal{Y}_{U_1} \right] 
    \frac{g_U}{\sqrt2 M_{U_1}}\,\beta_L^{ij} \,, \notag\\
   \frac{1}{\Lambda}\,C_{\bar Q_L G U_1 L_L}^{(2)\,ij}(\Lambda,M_{U_1},\mu_h,u) 
   &= - \left( 1 - \kappa_U \right) \frac{g_U}{\sqrt2 M_{U_1}}\,\beta_L^{ij} \,, \notag\\
   \frac{1}{\Lambda}\,C_{\bar Q_L W U_1 L_L}^{(2)\,ij}(\Lambda,M_{U_1},\mu_h,u) 
   &= \frac{g_U}{\sqrt2 M_{U_1}}\,\beta_L^{ij} \,, \notag\\
   \frac{1}{\Lambda}\,C_{\bar Q_L B U_1 L_L}^{(2)\,ij}(\Lambda,M_{U_1},\mu_h,u) 
   &= \left[ \mathcal{Y}_{Q_L} - \left( 2 - \tilde\kappa_U \right) \mathcal{Y}_{U_1} \right] 
    \frac{g_U}{\sqrt2 M_{U_1}}\,\beta_L^{ij} \,, \notag\\
   \frac{1}{\Lambda}\,C_{\bar d_R G U_1\ell_R}^{(1)\,ij}(\Lambda,M_{U_1},\mu_h,u)  
   &= - \left( 2 - \kappa_U \right) \frac{g_U}{\sqrt2 M_{U_1}}\,\beta_R^{ij} \,, \\   
   \frac{1}{\Lambda}\,C_{\bar d_R B U_1\ell_R}^{(1)\,ij}(\Lambda,M_{U_1},\mu_h,u)  
   &= \left[ \mathcal{Y}_{\ell_R} - \left( 2 - \tilde\kappa_U \right) \mathcal{Y}_{U_1} \right] 
    \frac{g_U}{\sqrt2 M_{U_1}}\,\beta_R^{ij} \,, \notag\\
   \frac{1}{\Lambda}\,C_{\bar d_R G U_1\ell_R}^{(2)\,ij}(\Lambda,M_{U_1},\mu_h,u)  
   &= - \left( 1 - \kappa_U \right) \frac{g_U}{\sqrt2 M_{U_1}}\,\beta_R^{ij} \,, \notag\\
   \frac{1}{\Lambda}\,C_{\bar d_R B U_1\ell_R}^{(2)\,ij}(\Lambda,M_{U_1},\mu_h,u)  
   &= \left[ \mathcal{Y}_{d_R} - \left( 2 - \tilde\kappa_U \right) \mathcal{Y}_{U_1} \right] 
    \frac{g_U}{\sqrt2 M_{U_1}}\,\beta_R^{ij} \,, \notag\\
   \frac{1}{\Lambda}\,C_{\bar u_R G U_1\nu_R}^{(1)\,ij}(\Lambda,M_{U_1},\mu_h,u) 
   &= - \left( 2 - \kappa_U \right) \frac{g_U^\nu}{M_{U_1}}\,\beta_\nu^{ij} \,, \notag\\
   \frac{1}{\Lambda}\,C_{\bar u_R B U_1\nu_R}^{(1)\,ij}(\Lambda,M_{U_1},\mu_h,u) 
   &= - \left( 2 - \tilde\kappa_U \right) \mathcal{Y}_{U_1} \frac{g_U^\nu}{M_{U_1}}\,\beta_\nu^{ij} \,, \notag\\
   \frac{1}{\Lambda}\,C_{\bar u_R G U_1\nu_R}^{(2)\,ij}(\Lambda,M_{U_1},\mu_h,u) 
   &= - \left( 1 - \kappa_U \right) \frac{g_U^\nu}{M_{U_1}}\,\beta_\nu^{ij} \,, \notag\\
   \frac{1}{\Lambda}\,C_{\bar u_R B U_1\nu_R}^{(2)\,ij}(\Lambda,M_{U_1},\mu_h,u) 
   &= \left[ \mathcal{Y}_{u_R} - \left( 2 - \tilde\kappa_U \right) \mathcal{Y}_{U_1} \right] 
    \frac{g_U^\nu}{M_{U_1}}\,\beta_\nu^{ij} \,. \notag
\end{align}
We use the symbol $\mathcal{Y}_f$ to denote the hypercharge values of the various SM fermion fields, and $\mathcal{Y}_{U_1}=\frac23$ is the hypercharge of the leptoquark. Note that the dependence on the momentum fraction $u$ cancels out in all diagrams. The remaining Wilson coefficients in (\ref{U1subLa}) vanish at tree level, i.e.\
\begin{equation}
   C_{\bar u_R A U_1\nu_R^c}^{(k)\,ij}(\Lambda,M_{U_1},\mu_h,u) 
   = C_{\bar u_R A U_1\nu_R^c}^{\prime\,(k)\,ij}(\Lambda,M_{U_1},\mu_h,u) 
   = 0 \,; \quad A=G,B \,,
\end{equation}
where $k=1,2$ in the last line. 

Focussing now on the operators containing a scalar doublet, defined in \eqref{LagrangianU1sub}, we find
\begin{equation}\label{matchingWU12nd}
\begin{aligned}
   \frac{1}{\Lambda}\,C_{\bar Q_L\Phi U_1\ell_R}^{(1)\,ij}(\Lambda,M_{U_1},\mu_h,u)   
   &= - \frac{g_U}{\sqrt2 M_{U_1}} \left( \bm{\beta}_L\bm{Y}_e \right)^{ij} , \\
   \frac{1}{\Lambda}\,C_{\bar Q_L\Phi U_1\ell_R}^{(2)\,ij}(\Lambda,M_{U_1},\mu_h,u)   
   &= - \frac{g_U}{\sqrt2 M_{U_1}} \left( \bm{Y}_d\,\bm{\beta}_R \right)^{ij} , \\
   \frac{1}{\Lambda}\,C_{\bar Q_L\tilde\Phi U_1\nu_R}^{(1)\,ij}(\Lambda,M_{U_1},\mu_h,u)   
   &= - \frac{g_U}{\sqrt2 M_{U_1}} \left( \bm{\beta}_L\bm{Y}_\nu \right)^{ij} , \\
   \frac{1}{\Lambda}\,C_{\bar Q_L\tilde\Phi U_1\nu_R}^{(2)\,ij}(\Lambda,M_{U_1},\mu_h,u)   
   &= - \frac{g_U^\nu}{M_{U_1}} \left( \bm{Y}_u\,\bm{\beta}_\nu \right)^{ij} , \\
   \frac{1}{\Lambda}\,C_{\bar d_R\Phi^\dagger U_1 L_L}^{(1)\,ij}(\Lambda,M_{U_1},\mu_h,u)   
   &= - \frac{g_U}{\sqrt2 M_{U_1}} \left( \bm{\beta}_R\bm{Y}_e^\dagger \right)^{ij} , \\
   \frac{1}{\Lambda}\,C_{\bar d_R\Phi^\dagger U_1 L_L}^{(2)\,ij}(\Lambda,M_{U_1},\mu_h,u)   
   &= - \frac{g_U}{\sqrt2 M_{U_1}} \left( \bm{Y}_d^\dagger\bm{\beta}_L \right)^{ij} , \\
   \frac{1}{\Lambda}\,C_{\bar u_R\tilde\Phi^\dagger U_1 L_L}^{(1)\,ij}(\Lambda,M_{U_1},\mu_h,u)   
   &= - \frac{g_U^\nu}{M_{U_1}} \left( \bm{\beta}_\nu\bm{Y}_\nu^\dagger \right)^{ij} , \\
   \frac{1}{\Lambda}\,C_{\bar u_R\tilde\Phi^\dagger U_1 L_L}^{(2)\,ij}(\Lambda,M_{U_1},\mu_h,u)   
   &= - \frac{g_U}{\sqrt2 M_{U_1}} \left( \bm{Y}_u^\dagger\bm{\beta}_L \right)^{ij} .
\end{aligned}
\end{equation}
The Wilson coefficients of the remaining operators vanish at tree level, i.e.\
\begin{equation}
\begin{aligned}
   C_{\bar Q_L\Phi U_1\ell_R}^{(0)\,ij}(\Lambda,M_{U_1},\mu_h)   
   &= 0 \,,, \\
   C_{\bar Q_L\tilde\Phi U_1\nu_R}^{(0)\,ij}(\Lambda,M_{U_1},\mu_h,u)   
   &= 0 \,, \\
   C_{\bar d_R\Phi^\dagger U_1 L_L}^{(0)\,ij}(\Lambda,M_{U_1},\mu_h,u)   
   &= 0 \,, \\
   C_{\bar u_R\tilde\Phi^\dagger U_1 L_L}^{(0)\,ij}(\Lambda,M_{U_1},\mu_h,u)   
   &= 0 \,, \\
   C_{\bar Q_L\tilde\Phi U_1\nu_R^c}^{(k)\,ij}(\Lambda,M_{U_1},\mu_h,u)  
   &= 0 \,, \\
   C_{\bar u_R\tilde\Phi U_1 L_L^c}^{(k)\,ij}(\Lambda,M_{U_1},\mu_h,u)  
   &= 0 \,, 
\end{aligned}
\end{equation}
where in the last two cases $k=0,1,2$.

The Wilson coefficients of the fermion-number violating operators $\mathcal{O}_{\bar Q_L\tilde\Phi U_1\nu_R^c}^{(k)\,ij}$ and $\mathcal{O}_{\bar u_R\tilde\Phi U_1 L_L^c}^{(k)\,ij}$ can be non-zero in UV models featuring additional new heavy particles. In the model containing a heavy right-handed neutrino with mass $M_\nu\gtrsim M_{U_1}$, we obtain
\begin{equation}
\begin{aligned}
   C_{\bar u_R\tilde\Phi U_1 L_L^c}^{(0)\,ij}(\Lambda,M_{U_1},\mu_h)
   &= C_{\bar u_R\tilde\Phi U_1 L_L^c}^{(2)\,ij}(\Lambda,M_{U_1},\mu_h,u)
    = \frac{g_U^\nu}{M_\nu} \left( \bm{\beta}_\nu\bm{Y}_\nu^T \right)^{ij} , \\
   C_{\bar u_R\tilde\Phi U_1 L_L^c}^{(1)\,ij}(\Lambda,M_{U_1},\mu_h,u)
   &= \frac{g_U^\nu M_\nu}{M_\nu^2-u M_{U_1}^2} \left( \bm{\beta}_\nu\bm{Y}_\nu^T \right)^{ij} .
\end{aligned}
\end{equation}

\section{Conclusions}
\label{conclusions}

In this work we have applied the SCET framework to present a detailed discussion of the decay rates of three well-motivated new particles not contained in the SM: the scalar leptoquarks $S_1$ and $S_3$, as well as the vector leptoquark $U_1$. These particles have been studied extensively in the context of the $B$-meson flavor anomalies. Several concrete models aiming to explain these anomalies include one or more of these leptoquarks, typically either $S_1$ and $S_3$, or $U_1$. It is therefore of particular interest to better understand the decay rates of these particles into SM particles. 

A consistent analysis of this problem based on the power counting $\lambda=v/\Lambda\ll 1$ (where $v$ denotes the electroweak scale and $\Lambda$ the New Physics scale set by the mass of the leptoquark) requires describing the leptoquark interactions with SM particles in terms of SCET operators. In this way, decay amplitudes can be systematically expanded in powers of $\lambda$ and large logarithmic corrections in $\lambda$ can be resummed using RG equations. In this paper, we have constructed in a model-independent way the most general operator basis for two-body (or, more generally, two-jet) decay processes at leading and subleading order in $\lambda$. For one of the leptoquarks, the scalar $S_1$, we have also constructed a basis of the leading-order three-jet operators, which can mediate interesting decay channels involving two fermions and a gauge or Higgs boson in the final state. A light right-handed neutrino is also allowed as a final-state particle, thus allowing for a minimal extension of the SM. Going to subleading order in the SCET expansion is motivated by the fact that this allows for new and rather different decay modes. In particular, for the scalar leptoquarks $S_1$ and $S_3$ all decays allowed at leading power violate fermion number by two units, where the decay modes allowed at subleading order conserve fermion number. In generic New Physics models both options are allowed.

We have solved the RG evolution equations for the SCET basis operators to resum the large (double and single) logarithmic corrections to the decay rate at leading order in RG-improved perturbation theory. We have presented numerical estimates of the resummation effects for some decays of strong phenomenological interest, finding that for all the three leptoquarks there is a significant effect coming from single logarithmic terms, which tend to compensate the suppression of the rates from Sudakov double logarithms. The decay rates would change by as much as 20\% if these single logarithmic terms were not properly included. 

Finally, we have derived the tree-level matching conditions for the Wilson coefficients of the two-jet SCET operators (at leading and subleading order in power counting) in concrete UV models for the three leptoquarks. With the example of a heavy right-handed neutrino added to the SM, we demonstrate explicitly that the scalar leptoquark $S_1$ can have both fermion-number violating and fermion-number conserving decay modes, which are described by leading and subleading SCET operators, respectively. Which decay modes are most relevant for phenomenology depends on the model parameters.

While previous applications of the SCET-BSM framework developed in \cite{Alte:2018nbn} have focused on heavy new particles that are singlets under the SM \cite{Alte:2019iug,Heiles:2020plj}, with this work we have extended the formalism to describe exotic heavy particles carrying non-trivial SM charges. In this case, the fields for the heavy resonances themselves must be defined in a heavy-particle effective theory. This work thus lays the theoretical foundations for further applications of the SCET-BSM framework to study heavy degrees of freedom not contained in the SM.

\subsubsection*{Acknowledgments}

We are grateful to Javier Fuentes-Mart\'in for useful discussions. M.N.~thanks Gino Isidori, the particle theory group at Zurich University and the Pauli Center for hospitality during a sabbatical stay. This research was supported by the Cluster of Excellence {\em Precision Physics, Fundamental Interactions and Structure of Matter\/} (PRISMA${}^+$ -- EXC~2118/1) within the German Excellence Strategy (project ID 39083149).

\newpage
\appendix

\renewcommand{\theequation}{A.\arabic{equation}}
\setcounter{equation}{0}

\section{Anomalous dimensions for three-jet operators}
\label{anomalousdim3jet}

It is not difficult to extract from the master formula \eqref{master} the anomalous dimensions of three-jet operators in the SCET formalism. We now illustrate this for the case of the scalar leptoquark $S_1$, for which the relevant operators have been defined in \eqref{ZandgammadecayS1}. Using momentum conservation, we can rewrite the argument of the second cusp logarithm in the master formula in the form 
\begin{equation}
   2v\cdot p_1 = \frac{1}{M_{S_1}} \left[ M_{S_1}^2 + m_1^2 - (M_{S_1} v-p_1)^2 \right]
   \approx \frac{M_{S_1}^2-m_{23}^2}{M_{S_1}} \,,
\end{equation}
and similarly for other scalar products. We obtain 
\begin{equation}
\begin{aligned}
   \bm{\Gamma}_{\bar d_R\tilde\Phi^\dagger L_L S_1} 
   &= - \frac{3}{4} \gamma^{(2)}_{\text{cusp}} \left( \ln\frac{\mu^2}{m_{\Phi L}^2} + i\pi \right)
    + \frac{1}{6} \gamma_{\text{cusp}}^{(1)} \ln\frac{m_{dL}^2}{m_{d\Phi}^2} \\
   &\quad - \left( \frac{4}{3} \gamma_{\text{cusp}}^{(3)} + \frac{1}{9} \gamma_{\text{cusp}}^{(1)} \right)
    \ln\frac{\mu M_{S_1}}{M_{S_1}-m_{\Phi L}^2}
    + \frac{1}{6} \gamma^{(1)}_{\text{cusp}} \ln\frac{M_{S_1}^2-m_{dL}^2}{M_{S_1}^2-m_{d\Phi}^2} \\
   &\quad + \gamma^{S_1} + \bm{\gamma}^{L_L} + \gamma^{\Phi} + \bm{\gamma}^{d_R} \,, \\
   \bm{\Gamma}_{\bar Q_L\Phi\nu_R S_1} 
   &= - \left( \frac{3}{4} \gamma_{\text{cusp}}^{(2)} + \frac{1}{12} \gamma_{\text{cusp}}^{(1)} \right)
    \left( \ln\frac{\mu^2}{m_{Q\Phi}^2} + i\pi \right) \\
   &\quad - \frac{4}{3} \gamma_{\text{cusp}}^{(3)} \ln\frac{\mu M_{S_1}}{M_{S_1}^2-m_{\Phi\nu}^2}
    + \frac{1}{18} \gamma_{\text{cusp}}^{(1)} \ln\frac{\mu M_{S_1}}{M_{S_1}^2-m_{\Phi\nu}^2}
    - \frac{1}{6} \gamma_{\text{cusp}}^{(1)} \ln\frac{\mu M_{S_1}}{M_{S_1}^2-m_{Q\nu}^2} \\
   &\quad +\gamma^ {S_1} + \gamma^{\Phi} + \bm{\gamma}^{Q_L} \,, \\
   \bm{\Gamma}_{\bar d_R B\nu_R S_1} 
   &= - \left( \frac{4}{3} \gamma_{\text{cusp}}^{(3)} + \frac{1}{9}\gamma_{\text{cusp}}^{(1)} \right)
    \ln\frac{\mu M_{S_1}}{M_{S_1}^2-m_{B\nu}^2} + \gamma^{S_1} + \gamma^B +\bm{\gamma}^{d_R} \,.
\end{aligned}
\end{equation}


\begin{thebibliography}{99}

\bibitem{Buchmuller:1986zs}
W.~Buchm\"uller, R.~R\"uckl and D.~Wyler,
%``Leptoquarks in Lepton - Quark Collisions,''
Phys. Lett. B \textbf{191}, 442-448 (1987)
[Erratum: Phys. Lett. B \textbf{448}, 320-320 (1999)].
%doi:10.1016/0370-2693(87)90637-X

\bibitem{Leurer:1993ap}
M.~Leurer,
%``New bounds on leptoquarks,''
Phys. Rev. Lett. \textbf{71}, 1324-1327 (1993)
%doi:10.1103/PhysRevLett.71.1324
[arXiv:hep-ph/9304211 [hep-ph]].

\bibitem{Davidson:1993qk}
S.~Davidson, D.~C.~Bailey and B.~A.~Campbell,
%``Model independent constraints on leptoquarks from rare processes,''
Z. Phys. C \textbf{61}, 613-644 (1994)
%doi:10.1007/BF01552629
[arXiv:hep-ph/9309310 [hep-ph]].

\bibitem{Plehn:1997az}
T.~Plehn, H.~Spiesberger, M.~Spira and P.~M.~Zerwas,
%``Formation and decay of scalar leptoquarks / squarks in e p collisions,''
Z. Phys. C \textbf{74}, 611-614 (1997)
%doi:10.1007/s002880050426
[arXiv:hep-ph/9703433 [hep-ph]].

\bibitem{Carpentier:2010ue}
M.~Carpentier and S.~Davidson,
%``Constraints on two-lepton, two quark operators,''
Eur. Phys. J. C \textbf{70}, 1071-1090 (2010)
%doi:10.1140/epjc/s10052-010-1482-4
[arXiv:1008.0280 [hep-ph]].

\bibitem{Dorsner:2017ufx}
I.~Dor\v{s}ner, S.~Fajfer, D.~A.~Faroughy and N.~Ko\v{s}nik,
%``The role of the $S_3$ GUT leptoquark in flavor universality and collider searches,''
JHEP \textbf{10}, 188 (2017)
%doi:10.1007/JHEP10(2017)188
[arXiv:1706.07779 [hep-ph]].

\bibitem{Pati:1974yy}
J.~C.~Pati and A.~Salam,
%``Lepton Number as the Fourth Color,''
Phys. Rev. D \textbf{10}, 275-289 (1974)
[Erratum: Phys. Rev. D \textbf{11}, 703-703 (1975)].
%doi:10.1103/PhysRevD.10.275

\bibitem{Georgi:1974sy}
H.~Georgi and S.~L.~Glashow,
%``Unity of All Elementary Particle Forces,''
Phys. Rev. Lett. \textbf{32}, 438-441 (1974).
%doi:10.1103/PhysRevLett.32.438

\bibitem{Georgi:1974yf}
H.~Georgi, H.~R.~Quinn and S.~Weinberg,
%``Hierarchy of Interactions in Unified Gauge Theories,''
Phys. Rev. Lett. \textbf{33}, 451-454 (1974).
%doi:10.1103/PhysRevLett.33.451

\bibitem{Fritzsch:1974nn}
H.~Fritzsch and P.~Minkowski,
%``Unified Interactions of Leptons and Hadrons,''
Annals Phys. \textbf{93}, 193-266 (1975).
%doi:10.1016/0003-4916(75)90211-0

\bibitem{BaBar:2012obs}
J.~P.~Lees \textit{et al.} [BaBar],
%``Evidence for an excess of $\bar{B} \to D^{(*)} \tau^-\bar{\nu}_\tau$ decays,''
Phys. Rev. Lett. \textbf{109}, 101802 (2012)
%doi:10.1103/PhysRevLett.109.101802
[arXiv:1205.5442 [hep-ex]].

\bibitem{BaBar:2013mob}
J.~P.~Lees \textit{et al.} [BaBar],
%``Measurement of an Excess of $\bar{B} \to D^{(*)}\tau^- \bar{\nu}_\tau$ Decays and Implications for Charged Higgs Bosons,''
Phys. Rev. D \textbf{88}, no.7, 072012 (2013)
%doi:10.1103/PhysRevD.88.072012
[arXiv:1303.0571 [hep-ex]].

\bibitem{LHCb:2015gmp}
R.~Aaij \textit{et al.} [LHCb],
%``Measurement of the ratio of branching fractions $\mathcal{B}(\bar{B}^0 \to D^{*+}\tau^{-}\bar{\nu}_{\tau})/\mathcal{B}(\bar{B}^0 \to D^{*+}\mu^{-}\bar{\nu}_{\mu})$,''
Phys. Rev. Lett. \textbf{115}, no.11, 111803 (2015)
[Erratum: Phys. Rev. Lett. \textbf{115}, no.15, 159901 (2015)]
%doi:10.1103/PhysRevLett.115.111803
[arXiv:1506.08614 [hep-ex]].

\bibitem{Belle:2015qfa}
M.~Huschle \textit{et al.} [Belle],
%``Measurement of the branching ratio of $\bar{B} \to D^{(\ast)} \tau^- \bar{\nu}_\tau$ relative to $\bar{B} \to D^{(\ast)} \ell^- \bar{\nu}_\ell$ decays with hadronic tagging at Belle,''
Phys. Rev. D \textbf{92}, no.7, 072014 (2015)
%doi:10.1103/PhysRevD.92.072014
[arXiv:1507.03233 [hep-ex]].

\bibitem{Belle:2016dyj}
S.~Hirose \textit{et al.} [Belle],
%``Measurement of the $\tau$ lepton polarization and $R(D^*)$ in the decay $\bar{B} \to D^* \tau^- \bar{\nu}_\tau$,''
Phys. Rev. Lett. \textbf{118}, no.21, 211801 (2017)
%doi:10.1103/PhysRevLett.118.211801
[arXiv:1612.00529 [hep-ex]].

\bibitem{LHCb:2017smo}
R.~Aaij \textit{et al.} [LHCb],
%``Measurement of the ratio of the $B^0 \to D^{*-} \tau^+ \nu_{\tau}$ and $B^0 \to D^{*-} \mu^+ \nu_{\mu}$ branching fractions using three-prong $\tau$-lepton decays,''
Phys. Rev. Lett. \textbf{120}, no.17, 171802 (2018)
%doi:10.1103/PhysRevLett.120.171802
[arXiv:1708.08856 [hep-ex]].

\bibitem{LHCb:2017rln}
R.~Aaij \textit{et al.} [LHCb],
%``Test of Lepton Flavor Universality by the measurement of the $B^0 \to D^{*-} \tau^+ \nu_{\tau}$ branching fraction using three-prong $\tau$ decays,''
Phys. Rev. D \textbf{97}, no.7, 072013 (2018)
%doi:10.1103/PhysRevD.97.072013
[arXiv:1711.02505 [hep-ex]].

\bibitem{LHCb:2014vgu}
R.~Aaij \textit{et al.} [LHCb],
%``Test of lepton universality using $B^{+}\rightarrow K^{+}\ell^{+}\ell^{-}$ decays,''
Phys. Rev. Lett. \textbf{113}, 151601 (2014)
%doi:10.1103/PhysRevLett.113.151601
[arXiv:1406.6482 [hep-ex]].

\bibitem{LHCb:2017avl}
R.~Aaij \textit{et al.} [LHCb],
%``Test of lepton universality with $B^{0} \rightarrow K^{*0}\ell^{+}\ell^{-}$ decays,''
JHEP \textbf{08}, 055 (2017)
%doi:10.1007/JHEP08(2017)055
[arXiv:1705.05802 [hep-ex]].

\bibitem{LHCb:2019hip}
R.~Aaij \textit{et al.} [LHCb],
%``Search for lepton-universality violation in $B^+\to K^+\ell^+\ell^-$ decays,''
Phys. Rev. Lett. \textbf{122}, no.19, 191801 (2019)
%doi:10.1103/PhysRevLett.122.191801
[arXiv:1903.09252 [hep-ex]].

\bibitem{LHCb:2021trn}
R.~Aaij \textit{et al.} [LHCb],
%``Test of lepton universality in beauty-quark decays,''
Nature Phys. \textbf{18}, no.3, 277-282 (2022)
%doi:10.1038/s41567-021-01478-8
[arXiv:2103.11769 [hep-ex]].

\bibitem{LHCb:2021lvy}
R.~Aaij \textit{et al.} [LHCb],
%``Tests of lepton universality using $B^0\to K^0_S \ell^+ \ell^-$ and $B^+\to K^{*+} \ell^+ \ell^-$ decays,''
Phys. Rev. Lett. \textbf{128}, no.19, 191802 (2022)
%doi:10.1103/PhysRevLett.128.191802
[arXiv:2110.09501 [hep-ex]].

\bibitem{Sakaki:2013bfa}
Y.~Sakaki, M.~Tanaka, A.~Tayduganov and R.~Watanabe,
%``Testing leptoquark models in $\bar B \to D^{(*)} \tau \bar\nu$,''
Phys. Rev. D \textbf{88}, no.9, 094012 (2013)
%doi:10.1103/PhysRevD.88.094012
[arXiv:1309.0301 [hep-ph]].

\bibitem{Hiller:2014yaa}
G.~Hiller and M.~Schmaltz,
%``$R_K$ and future $b \to s \ell \ell$ physics beyond the standard model opportunities,''
Phys. Rev. D \textbf{90}, 054014 (2014)
%doi:10.1103/PhysRevD.90.054014
[arXiv:1408.1627 [hep-ph]].

\bibitem{Buras:2014fpa}
A.~J.~Buras, J.~Girrbach-Noe, C.~Niehoff and D.~M.~Straub,
%``$ B\to {K}^{\left(\ast \right)}\nu \overline{\nu} $ decays in the Standard Model and beyond,''
JHEP \textbf{02}, 184 (2015)
%doi:10.1007/JHEP02(2015)184
[arXiv:1409.4557 [hep-ph]].

\bibitem{Gripaios:2014tna}
B.~Gripaios, M.~Nardecchia and S.~A.~Renner,
%``Composite leptoquarks and anomalies in $B$-meson decays,''
JHEP \textbf{05}, 006 (2015)
%doi:10.1007/JHEP05(2015)006
[arXiv:1412.1791 [hep-ph]].

\bibitem{deMedeirosVarzielas:2015yxm}
I.~de Medeiros Varzielas and G.~Hiller,
%``Clues for flavor from rare lepton and quark decays,''
JHEP \textbf{06}, 072 (2015)
%doi:10.1007/JHEP06(2015)072
[arXiv:1503.01084 [hep-ph]].

\bibitem{Bauer:2015knc}
M.~Bauer and M.~Neubert,
%``Minimal Leptoquark Explanation for the $R_{D^{(*)}}$ , $R_K$ , and $(g-2)_\mu$ Anomalies,''
Phys. Rev. Lett. \textbf{116}, no.14, 141802 (2016)
%doi:10.1103/PhysRevLett.116.141802
[arXiv:1511.01900 [hep-ph]].

\bibitem{Barbieri:2015yvd}
R.~Barbieri, G.~Isidori, A.~Pattori and F.~Senia,
%``Anomalies in $B$-decays and $U(2)$ flavour symmetry,''
Eur. Phys. J. C \textbf{76}, no.2, 67 (2016)
%doi:10.1140/epjc/s10052-016-3905-3
[arXiv:1512.01560 [hep-ph]].

\bibitem{Duraisamy:2016gsd}
M.~Duraisamy, S.~Sahoo and R.~Mohanta,
%``Rare semileptonic $B \to K(\pi)l_i^- l_j^+$ decay in a vector leptoquark model,''
Phys. Rev. D \textbf{95}, no.3, 035022 (2017)
%doi:10.1103/PhysRevD.95.035022
[arXiv:1610.00902 [hep-ph]].

\bibitem{Cox:2016epl}
P.~Cox, A.~Kusenko, O.~Sumensari and T.~T.~Yanagida,
%``SU(5) Unification with TeV-scale Leptoquarks,''
JHEP \textbf{03}, 035 (2017)
%doi:10.1007/JHEP03(2017)035
[arXiv:1612.03923 [hep-ph]].

\bibitem{Crivellin:2017zlb}
A.~Crivellin, D.~M\"uller and T.~Ota,
%``Simultaneous explanation of R(D$^{(∗)}$) and b\textrightarrow{}s\ensuremath{\mu}$^{+}$ \ensuremath{\mu}$^{−}$: the last scalar leptoquarks standing,''
JHEP \textbf{09}, 040 (2017)
%doi:10.1007/JHEP09(2017)040
[arXiv:1703.09226 [hep-ph]].

\bibitem{Hiller:2017bzc}
G.~Hiller and I.~Nisandzic,
%``$R_K$ and $R_{K^{\ast}}$ beyond the standard model,''
Phys. Rev. D \textbf{96}, no.3, 035003 (2017)
%doi:10.1103/PhysRevD.96.035003
[arXiv:1704.05444 [hep-ph]].

\bibitem{Cai:2017wry}
Y.~Cai, J.~Gargalionis, M.~A.~Schmidt and R.~R.~Volkas,
%``Reconsidering the One Leptoquark solution: flavor anomalies and neutrino mass,''
JHEP \textbf{10}, 047 (2017)
%doi:10.1007/JHEP10(2017)047
[arXiv:1704.05849 [hep-ph]].

\bibitem{Buttazzo:2017ixm}
D.~Buttazzo, A.~Greljo, G.~Isidori and D.~Marzocca,
%``B-physics anomalies: a guide to combined explanations,''
JHEP \textbf{11}, 044 (2017)
%doi:10.1007/JHEP11(2017)044
[arXiv:1706.07808 [hep-ph]].

\bibitem{DiLuzio:2017vat}
L.~Di Luzio, A.~Greljo and M.~Nardecchia,
%``Gauge leptoquark as the origin of B-physics anomalies,''
Phys. Rev. D \textbf{96}, no.11, 115011 (2017)
%doi:10.1103/PhysRevD.96.115011
[arXiv:1708.08450 [hep-ph]].

\bibitem{Calibbi:2017qbu}
L.~Calibbi, A.~Crivellin and T.~Li,
%``Model of vector leptoquarks in view of the $B$-physics anomalies,''
Phys. Rev. D \textbf{98}, no.11, 115002 (2018)
%doi:10.1103/PhysRevD.98.115002
[arXiv:1709.00692 [hep-ph]].

\bibitem{Bordone:2017bld}
M.~Bordone, C.~Cornella, J.~Fuentes-Mart\'\i{}n and G.~Isidori,
%``A three-site gauge model for flavor hierarchies and flavor anomalies,''
Phys. Lett. B \textbf{779}, 317-323 (2018)
%doi:10.1016/j.physletb.2018.02.011
[arXiv:1712.01368 [hep-ph]].

\bibitem{Smirnov:2018ske}
A.~D.~Smirnov,
%``Vector leptoquark mass limits and branching ratios of $ K_L^0, B^0, B_s \to l^+_i l^-_j $ decays with account of fermion mixing in leptoquark currents,''
Mod. Phys. Lett. A \textbf{33}, 1850019 (2018)
%doi:10.1142/S0217732318500190
[arXiv:1801.02895 [hep-ph]].

\bibitem{Blanke:2018sro}
M.~Blanke and A.~Crivellin,
%``$B$ Meson Anomalies in a Pati-Salam Model within the Randall-Sundrum Background,''
Phys. Rev. Lett. \textbf{121}, no.1, 011801 (2018)
%doi:10.1103/PhysRevLett.121.011801
[arXiv:1801.07256 [hep-ph]].

\bibitem{Greljo:2018tuh}
A.~Greljo and B.~A.~Stefanek,
%``Third family quark\textendash{}lepton unification at the TeV scale,''
Phys. Lett. B \textbf{782}, 131-138 (2018)
%doi:10.1016/j.physletb.2018.05.033
[arXiv:1802.04274 [hep-ph]].

\bibitem{Bordone:2018nbg}
M.~Bordone, C.~Cornella, J.~Fuentes-Mart\'\i{}n and G.~Isidori,
%``Low-energy signatures of the $\mathrm{PS}^3$ model: from $B$-physics anomalies to LFV,''
JHEP \textbf{10}, 148 (2018)
%doi:10.1007/JHEP10(2018)148
[arXiv:1805.09328 [hep-ph]].

\bibitem{Angelescu:2018tyl}
A.~Angelescu, D.~Be\v{c}irevi\'c, D.~A.~Faroughy and O.~Sumensari,
%``Closing the window on single leptoquark solutions to the $B$-physics anomalies,''
JHEP \textbf{10}, 183 (2018)
%doi:10.1007/JHEP10(2018)183
[arXiv:1808.08179 [hep-ph]].

\bibitem{Cornella:2019hct}
C.~Cornella, J.~Fuentes-Mart\'\i{}n and G.~Isidori,
%``Revisiting the vector leptoquark explanation of the B-physics anomalies,''
JHEP \textbf{07}, 168 (2019)
%doi:10.1007/JHEP07(2019)168
[arXiv:1903.11517 [hep-ph]].

\bibitem{Bigaran:2019bqv}
I.~Bigaran, J.~Gargalionis and R.~R.~Volkas,
%``A near-minimal leptoquark model for reconciling flavour anomalies and generating radiative neutrino masses,''
JHEP \textbf{10}, 106 (2019)
%doi:10.1007/JHEP10(2019)106
[arXiv:1906.01870 [hep-ph]].

\bibitem{Cornella:2021sby}
C.~Cornella, D.~A.~Faroughy, J.~Fuentes-Mart\'\i{}n, G.~Isidori and M.~Neubert,
%``Reading the footprints of the B-meson flavor anomalies,''
JHEP \textbf{08}, 050 (2021)
%doi:10.1007/JHEP08(2021)050
[arXiv:2103.16558 [hep-ph]].

\bibitem{Chakraverty:2001yg}
D.~Chakraverty, D.~Choudhury and A.~Datta,
%``A Nonsupersymmetric resolution of the anomalous muon magnetic moment,''
Phys. Lett. B \textbf{506}, 103-108 (2001)
%doi:10.1016/S0370-2693(01)00419-1
[arXiv:hep-ph/0102180 [hep-ph]].

\bibitem{Cheung:2001ip}
K.~m.~Cheung,
%``Muon anomalous magnetic moment and leptoquark solutions,''
Phys. Rev. D \textbf{64}, 033001 (2001)
%doi:10.1103/PhysRevD.64.033001
[arXiv:hep-ph/0102238 [hep-ph]].

\bibitem{CMS:2018qqq}
A.~M.~Sirunyan \textit{et al.} [CMS],
%``Constraints on models of scalar and vector leptoquarks decaying to a quark and a neutrino at $\sqrt{s}=$ 13 TeV,''
Phys. Rev. D \textbf{98}, no.3, 032005 (2018)
%doi:10.1103/PhysRevD.98.032005
[arXiv:1805.10228 [hep-ex]].

\bibitem{ATLAS:2020dsk}
G.~Aad \textit{et al.} [ATLAS],
%``Search for pairs of scalar leptoquarks decaying into quarks and electrons or muons in $ \sqrt{s} $ = 13 TeV $pp$ collisions with the ATLAS detector,''
JHEP \textbf{10}, 112 (2020)
%doi:10.1007/JHEP10(2020)112
[arXiv:2006.05872 [hep-ex]].

\bibitem{ATLAS:2020xov}
G.~Aad \textit{et al.} [ATLAS],
%``Search for pair production of scalar leptoquarks decaying into first- or second-generation leptons and top quarks in proton\textendash{}proton collisions at $\sqrt{s}$ = 13 TeV with the ATLAS detector,''
Eur. Phys. J. C \textbf{81}, no.4, 313 (2021)
%doi:10.1140/epjc/s10052-021-09009-8
[arXiv:2010.02098 [hep-ex]].

\bibitem{Bauer:2000yr}
C.~W.~Bauer, S.~Fleming, D.~Pirjol and I.~W.~Stewart,
%``An Effective field theory for collinear and soft gluons: Heavy to light decays,''
Phys. Rev. D \textbf{63}, 114020 (2001)
%doi:10.1103/PhysRevD.63.114020
[arXiv:hep-ph/0011336 [hep-ph]].

\bibitem{Bauer:2001yt}
C.~W.~Bauer, D.~Pirjol and I.~W.~Stewart,
%``Soft collinear factorization in effective field theory,''
Phys. Rev. D \textbf{65}, 054022 (2002)
%doi:10.1103/PhysRevD.65.054022
[arXiv:hep-ph/0109045 [hep-ph]].

\bibitem{Bauer:2002nz}
C.~W.~Bauer, S.~Fleming, D.~Pirjol, I.~Z.~Rothstein and I.~W.~Stewart,
%``Hard scattering factorization from effective field theory,''
Phys. Rev. D \textbf{66}, 014017 (2002)
%doi:10.1103/PhysRevD.66.014017
[arXiv:hep-ph/0202088 [hep-ph]].

\bibitem{Beneke:2002ph}
M.~Beneke, A.~P.~Chapovsky, M.~Diehl and T.~Feldmann,
%``Soft collinear effective theory and heavy to light currents beyond leading power,''
Nucl. Phys. B \textbf{643}, 431-476 (2002)
%doi:10.1016/S0550-3213(02)00687-9
[arXiv:hep-ph/0206152 [hep-ph]].

\bibitem{Becher:2014oda}
T.~Becher, A.~Broggio and A.~Ferroglia,
%``Introduction to Soft-Collinear Effective Theory,''
Lect. Notes Phys. \textbf{896}, pp.1-206 (2015),
Springer (2015),
%doi:10.1007/978-3-319-14848-9
[arXiv:1410.1892 [hep-ph]].

\bibitem{Alte:2018nbn}
S.~Alte, M.~K\"onig and M.~Neubert,
%``Effective Field Theory after a New-Physics Discovery,''
JHEP \textbf{08}, 095 (2018)
[Erratum: JHEP \textbf{04}, 009 (2021)]
%doi:10.1007/JHEP08(2018)095
[arXiv:1806.01278 [hep-ph]].

\bibitem{Alte:2019iug}
S.~Alte, M.~K\"onig and M.~Neubert,
%``Effective Theory for a Heavy Scalar Coupled to the SM via Vector-Like Quarks,''
Eur. Phys. J. C \textbf{79}, no.4, 352 (2019)
%doi:10.1140/epjc/s10052-019-6867-4
[arXiv:1902.04593 [hep-ph]].

\bibitem{Heiles:2020plj}
M.~Heiles, M.~K\"onig and M.~Neubert,
%``Effective Field Theory for Heavy Vector Resonances Coupled to the Standard Model,''
JHEP \textbf{02}, 204 (2021)
%doi:10.1007/JHEP02(2021)204
[arXiv:2011.08205 [hep-ph]].

\bibitem{Mohapatra:1980qe}
R.~N.~Mohapatra and R.~E.~Marshak,
%``Local B-L Symmetry of Electroweak Interactions, Majorana Neutrinos and Neutron Oscillations,''
Phys. Rev. Lett. \textbf{44}, 1316-1319 (1980)
[Erratum: Phys. Rev. Lett. \textbf{44}, 1643 (1980)].
%doi:10.1103/PhysRevLett.44.1316

\bibitem{Wetterich:1981bx}
C.~Wetterich,
%``Neutrino Masses and the Scale of B-L Violation,''
Nucl. Phys. B \textbf{187}, 343-375 (1981).
%doi:10.1016/0550-3213(81)90279-0

\bibitem{Buchmuller:1991ce}
W.~Buchm\"uller, C.~Greub and P.~Minkowski,
%``Neutrino masses, neutral vector bosons and the scale of B-L breaking,''
Phys. Lett. B \textbf{267}, 395-399 (1991).
%doi:10.1016/0370-2693(91)90952-M

\bibitem{Buchmuller:1992qc}
W.~Buchm\"uller and T.~Yanagida,
%``Baryogenesis and the scale of B-L breaking,''
Phys. Lett. B \textbf{302}, 240-244 (1993).
%doi:10.1016/0370-2693(93)90391-T

\bibitem{Khalil:2006yi}
S.~Khalil,
%``Low scale $B$ - L extension of the Standard Model at the LHC,''
J. Phys. G \textbf{35}, 055001 (2008)
%doi:10.1088/0954-3899/35/5/055001
[arXiv:hep-ph/0611205 [hep-ph]].

\bibitem{Abbas:2007ag}
M.~Abbas and S.~Khalil,
%``Neutrino masses, mixing and leptogenesis in TeV scale $B$ - L extension of the standard model,''
JHEP \textbf{04}, 056 (2008)
%doi:10.1088/1126-6708/2008/04/056
[arXiv:0707.0841 [hep-ph]].

\bibitem{Eichten:1989zv}
E.~Eichten and B.~R.~Hill,
%``An Effective Field Theory for the Calculation of Matrix Elements Involving Heavy Quarks,''
Phys. Lett. B \textbf{234}, 511-516 (1990).
%doi:10.1016/0370-2693(90)92049-O

\bibitem{Georgi:1990um}
H.~Georgi,
%``An Effective Field Theory for Heavy Quarks at Low-energies,''
Phys. Lett. B \textbf{240}, 447-450 (1990).
%doi:10.1016/0370-2693(90)91128-X

\bibitem{Neubert:1993mb}
M.~Neubert,
%``Heavy quark symmetry,''
Phys. Rept. \textbf{245}, 259-396 (1994)
%doi:10.1016/0370-1573(94)90091-4
[arXiv:hep-ph/9306320 [hep-ph]].

\bibitem{Manohar:2002fd}
A.~V.~Manohar, T.~Mehen, D.~Pirjol and I.~W.~Stewart,
%``Reparameterization invariance for collinear operators,''
Phys. Lett. B \textbf{539}, 59-66 (2002)
%doi:10.1016/S0370-2693(02)02029-4
[arXiv:hep-ph/0204229 [hep-ph]].

\bibitem{Hill:2002vw}
R.~J.~Hill and M.~Neubert,
%``Spectator interactions in soft collinear effective theory,''
Nucl. Phys. B \textbf{657}, 229-256 (2003)
%doi:10.1016/S0550-3213(03)00116-0
[arXiv:hep-ph/0211018 [hep-ph]].

\bibitem{Marcantonini:2008qn}
C.~Marcantonini and I.~W.~Stewart,
%``Reparameterization Invariant Collinear Operators,''
Phys. Rev. D \textbf{79}, 065028 (2009)
%doi:10.1103/PhysRevD.79.065028
[arXiv:0809.1093 [hep-ph]].

\bibitem{Bajc:2005zf}
B.~Bajc, A.~Melfo, G.~Senjanovic and F.~Vissani,
%``Yukawa sector in non-supersymmetric renormalizable SO(10),''
Phys. Rev. D \textbf{73}, 055001 (2006)
%doi:10.1103/PhysRevD.73.055001
[arXiv:hep-ph/0510139 [hep-ph]].

\bibitem{Catani:1996vz}
S.~Catani and M.~H.~Seymour,
%``A General algorithm for calculating jet cross-sections in NLO QCD,''
Nucl. Phys. B \textbf{485}, 291-419 (1997)
[Erratum: Nucl. Phys. B \textbf{510}, 503-504 (1998)]
%doi:10.1016/S0550-3213(96)00589-5
[arXiv:hep-ph/9605323 [hep-ph]].

\bibitem{Becher:2009qa}
T.~Becher and M.~Neubert,
%``On the Structure of Infrared Singularities of Gauge-Theory Amplitudes,''
JHEP \textbf{06}, 081 (2009)
[Erratum: JHEP \textbf{11}, 024 (2013)]
%doi:10.1088/1126-6708/2009/06/081
[arXiv:0903.1126 [hep-ph]].

\bibitem{Becher:2009kw}
T.~Becher and M.~Neubert,
%``Infrared singularities of QCD amplitudes with massive partons,''
Phys. Rev. D \textbf{79}, 125004 (2009)
[Erratum: Phys. Rev. D \textbf{80}, 109901 (2009)]
%doi:10.1103/PhysRevD.79.125004
[arXiv:0904.1021 [hep-ph]].

\bibitem{Korchemsky:1987wg}
G.~P.~Korchemsky and A.~V.~Radyushkin,
%``Renormalization of the Wilson Loops Beyond the Leading Order,''
Nucl. Phys. B \textbf{283}, 342-364 (1987).
%doi:10.1016/0550-3213(87)90277-X

\bibitem{Korchemskaya:1992je}
I.~A.~Korchemskaya and G.~P.~Korchemsky,
%``On lightlike Wilson loops,''
Phys. Lett. B \textbf{287}, 169-175 (1992).
%doi:10.1016/0370-2693(92)91895-G

\bibitem{Korchemskaya:1994qp}
I.~A.~Korchemskaya and G.~P.~Korchemsky,
%``High-energy scattering in QCD and cross singularities of Wilson loops,''
Nucl. Phys. B \textbf{437}, 127-162 (1995)
%doi:10.1016/0550-3213(94)00553-Q
[arXiv:hep-ph/9409446 [hep-ph]].

\bibitem{Moch:2004pa}
S.~Moch, J.~A.~M.~Vermaseren and A.~Vogt,
%``The Three loop splitting functions in QCD: The Nonsinglet case,''
Nucl. Phys. B \textbf{688}, 101-134 (2004)
%doi:10.1016/j.nuclphysb.2004.03.030
[arXiv:hep-ph/0403192 [hep-ph]].

\bibitem{Jantzen:2005az}
B.~Jantzen, J.~H.~K\"uhn, A.~A.~Penin and V.~A.~Smirnov,
%``Two-loop electroweak logarithms in four-fermion processes at high energy,''
Nucl. Phys. B \textbf{731}, 188-212 (2005)
[Erratum: Nucl. Phys. B \textbf{752}, 327-328 (2006)]
%doi:10.1016/j.nuclphysb.2005.10.010
[arXiv:hep-ph/0509157 [hep-ph]].

\bibitem{Grzadkowski:1987tf}
B.~Grzadkowski and M.~Lindner,
%``Nonlinear Evolution of Yukawa Couplings,''
Phys. Lett. B \textbf{193}, 71 (1987).
%doi:10.1016/0370-2693(87)90458-8

\bibitem{DiLuzio:2018zxy}
L.~Di Luzio, J.~Fuentes-Martin, A.~Greljo, M.~Nardecchia and S.~Renner,
%``Maximal Flavour Violation: a Cabibbo mechanism for leptoquarks,''
JHEP \textbf{11}, 081 (2018)
%doi:10.1007/JHEP11(2018)081
[arXiv:1808.00942 [hep-ph]].

\bibitem{Baker:2019sli}
M.~J.~Baker, J.~Fuentes-Mart\'\i{}n, G.~Isidori and M.~K\"onig,
%``High- $p_T$ signatures in vector\textendash{}leptoquark models,''
Eur. Phys. J. C \textbf{79}, no.4, 334 (2019)
%doi:10.1140/epjc/s10052-019-6853-x
[arXiv:1901.10480 [hep-ph]].
  
\end{thebibliography}
\end{document}